\documentclass[11pt]{article}
\usepackage{bbm}
\usepackage{epsfig}
\usepackage{array}
\usepackage{float}
\usepackage{dsfont}
\usepackage{amstext}
\usepackage{rotating}
\usepackage{a4}
\usepackage{a4wide}
\usepackage{cite}
\usepackage{multirow}

\def\be{\begin{equation}}
\def\ee{\end{equation}}
\def\gs{\mathrel{
   \rlap{\raise 0.511ex \hbox{$>$}}{\lower 0.511ex \hbox{$\sim$}}}}
\def\ls{\mathrel{
   \rlap{\raise 0.511ex \hbox{$<$}}{\lower 0.511ex \hbox{$\sim$}}}}

\newcommand{\ba}{\begin{array}{c}}
\newcommand{\baz}{\begin{array}{cc}}
\newcommand{\barrr}{\begin{array}{rrr}}
\newcommand{\bad}{\begin{array}{ccc}}
\newcommand{\bav}{\begin{array}{cccc}}
\newcommand{\baf}{\begin{array}{ccccc}}
\newcommand{\bea}{\begin{equation} \begin{array}{c}}
\newcommand{\eea}{ \end{array} \end{equation}}
\newcommand{\ea}{\end{array}}
\newcommand{\D}{\displaystyle}
\newcommand{\dms}{\mbox{$\Delta m^2_{\odot}$}}
\newcommand{\dma}{\mbox{$\Delta m^2_{\rm A}$}}
\newcommand{\meff}{\mbox{$\langle m \rangle$}}

\newcommand{\beqa}{\begin{eqnarray}}
\newcommand{\eeqa}{\end{eqnarray}}
\newcommand{\dmsq}{\Delta m^2}

\newcommand{\gsim}{\raise0.3ex\hbox{$\;>$\kern-0.75em\raise-1.1ex\hbox{$\sim\;$}}} 
\newcommand{\lsim}{\raise0.3ex\hbox{$\;<$\kern-0.75em\raise-1.1ex\hbox{$\sim\;$}}}
\newcommand{\betabeta}{\mbox{$(\beta\beta)_{0\nu}$}}

\begin{document}

\title{\vspace{-1.6cm}
\hfill {\small SISSA 40/2009/EP}\\[-0.1in]
\hfill {\small TIFR/TH/09-20}\\[-0.1in] 
\hfill {\small arXiv: 0907.2869 [hep-ph]}
\vskip 0.4cm
\bf 
Large $|U_{e3}|$ and Tri-bimaximal Mixing}
\author{
Srubabati Goswami$^{a}$\thanks{On lien from 
Harish--Chandra Research Institute, Chhatnag Road, Jhunsi, 
Allahabad 211 019, India.
email: \tt sruba@prl.res.in}\mbox{ },~~
Serguey T.~Petcov$^{b}$\thanks{Also at: Institute of
Nuclear Research and Nuclear Energy,
Bulgarian Academy of Sciences, 1784 Sofia, Bulgaria.
email: \tt petcov@sissa.it}\mbox{ },~~
Shamayita Ray$^{c}$\thanks{
email: \tt shamayitar@theory.tifr.res.in}\mbox{ },~~
Werner Rodejohann$^{d}$\thanks{email: 
\tt werner.rodejohann@mpi-hd.mpg.de}
\\ \\
{\normalsize \it $^a$Physical Research Laboratory,}\\
{\normalsize \it Ahmedabad 380 009, India}\\ \\
{\normalsize \it $^b$SISSA and INFN Sezione di Trieste,}\\
{\normalsize \it Via Beirut 2--4, I--34014 Trieste, Italy}\\
{\normalsize and}\\
{\normalsize \it Institute for the Physics of Mathematics of the Universe,}\\
{\normalsize \it 5-1-5 Kashiwa-no-Ha, Kashiwa Shi, Chiba 277-8568, Japan} \\ \\ 
{\normalsize \it $^c$Tata Institute of Fundamental Research,}\\
{\normalsize \it Homi Bhabha Road, Mumbai 400005, India}\\ \\
{\normalsize \it$^d$Max--Planck--Institut f\"ur Kernphysik,}\\
{\normalsize \it  Postfach 103980, D--69029 Heidelberg, Germany} 
}
\date{} 
\maketitle 
\thispagestyle{empty}
\vspace{-0.8cm}
\begin{abstract}
\noindent  
We investigate in a model-independent way to what extent one can perturb 
tri-bimaximal mixing in order to generate a 
sizable value of $|U_{e3}|$, while at the same time keeping   
solar neutrino mixing near its measured value, which is close to 
$\sin^2 \theta_{12} = \frac 13$. Three straightforward breaking
mechanisms to generate $|U_{e3}| \simeq 0.1$ are considered. 
For charged lepton corrections, the suppression of a sizable 
contribution to $\sin^2 \theta_{12}$ can be
achieved if CP violation in neutrino oscillations is almost 
maximal. 
Generation of  the 
indicated value of $|U_{e3}| \simeq 0.1$ 
through  renormalization group corrections
requires the neutrinos to be quasi-degenerate in mass. 
The consistency with the allowed range 
of $\sin^2\theta_{12}$ together with large running of 
$|U_{e3}|$ forces one of the Majorana phases to be close to $\pi$.     
This implies large cancellations in the effective Majorana mass 
governing neutrino-less double beta ($\betabeta$-)decay, 
constraining it to lie near its minimum allowed value of 
$m_0 \, \cos 2 \theta_{12}$, where $m_0 \gs 0.1$ eV. 
Finally, explicit breaking of the neutrino mass matrix in
the inverted hierarchical and quasi-degenerate neutrino mass spectrum 
cases is similarly correlated with
the $\betabeta$-decay effective Majorana mass, 
although to a lesser extent. 
The implied values for the atmospheric neutrino 
mixing angle $\theta_{23}$ are given in all cases.

\end{abstract}

\newpage

\section{Introduction}
It is a remarkable achievement of experimental 
neutrino physics to have identified the leading form 
of lepton mixing, or 
Pontecorvo-Maki-Nakagawa-Sakata (PMNS), mixing matrix 
\cite{BPont57} $U$:  
\be \label{eq:UTBM}
U \simeq U_{\rm TBM} \, P \, , \mbox{ where }  
U_{\rm TBM} = \left( 
\bad 
\sqrt{\frac 23} & \sqrt{\frac 13} & 0 \\
-\sqrt{\frac 16} &  \sqrt{\frac 13} &  -\sqrt{\frac 12} \\ 
-\sqrt{\frac 16} &  \sqrt{\frac 13} &  \sqrt{\frac 12}
\ea
\right) 
\ee
%
and $P = {\rm diag}(1, e^{i \alpha_2/2}, e^{i \alpha_3/2})$ 
contains the Majorana phases \cite{BHP80,SchVal80}. 
The above matrix Eq.~(\ref{eq:UTBM}) defines tri-bimaximal 
mixing (TBM) \cite{tri}: 
\be \label{eq:obsTBM}
\sin^2 \theta_{12} = \frac 13 ~,~~\sin^2 \theta_{23} = \frac 12 ~,~~
U_{e3} = 0\,.
\ee
%
Currently, the  values of 
$\sin^2 \theta_{12}$ and $\sin^2 \theta_{23}$
obtained from global fits of the 
neutrino oscillation data  
are indeed very close to those predicted by TBM \cite{bari}: 
\begin{equation}
\begin{array}{rcl}
\sin^2 \theta_{23} &=& 0.466^{+0.073,\,0.178}_{-0.058,\,0.135} 
\,,\\[0.24cm]
\sin^2 \theta_{12} &=& 0.312^{+0.019,\,0.063}_{-0.018,\,0.049} \,.\\
\end{array}
\label{eq:data}
\end{equation}  
%
Here we have given the best-fit values as well as 
the $1\sigma$ and $3\sigma$ ranges 
(see also \cite{concha,india,thomas}). 

Obviously, even if Nature has chosen TBM \footnote{The experimental 
results are so close to TBM that parameterizations of the PMNS 
matrix with TBM as the starting point have been proposed 
\cite{triM}.} as the lepton 
mixing scheme, one expects  deviations from it 
on very general grounds. 
Straightforward examples are charged lepton corrections 
(i.e., corrections stemming from the diagonalization 
of the charged lepton mass matrix), 
renormalization effects, or explicit breaking in the neutrino 
mass matrix giving rise to TBM. 

 Interestingly, in what regards the third 
mixing angle $\theta_{13}$, a weak indication 
towards a non-vanishing value has recently emerged from a 
combination of two independent hints in solar, reactor 
and atmospheric neutrino data. 
Reference \cite{bari} quotes the following best-fit value 
and $1\sigma$ range:  
\begin{equation}
\sin^2 \theta_{13} = 0.016 \pm 0.010\, , 
\label{eq:bari}
\end{equation}
%
or $|U_{e3}| = \sin \theta_{13} = 0.126_{-0.049}^{+0.035}$, or 
$\theta_{13} = \left(7.3_{-2.8}^{+2.0} \right)^\circ$.  
Vanishing $\theta_{13}$ is thus disfavored at $1.6\sigma$. Similar 
values and ranges have been found in other, independent 
analyses \cite{other0}. We note that the hint in the atmospheric data has been 
questioned \cite{MS}, but that the recent MINOS data show an excess of electron 
events \cite{MINOS}, which may be interpreted \cite{MINOS_bari} as another 
hint for a non-zero $\theta_{13}$. 

  The allowed ranges for the mass-squared differences from the 
current global fit performed in \cite{bari} are 
\begin{equation}
\begin{array}{rcl}
\dms &=& 7.67^{+0.16,\,0.52}_{-0.19,\,0.53} \times 10^{-5} {\mathrm {eV}}^2
\,,\\[0.24cm]
|\dma| &=& 2.39^{+0.11,\,0.42}_{-0.08,\,0.33} \times 10^{-3} {\mathrm {eV}}^2 \,.\\
\end{array}
\label{eq:massdata}
\end{equation}
%
Note that the sign of $|\dma| \simeq |\Delta m^2_{31}| \simeq 
|\Delta m^2_{32}|$, i.e., the ordering of neutrino masses 
is still not known. Regarding the neutrino mass scale, 
there are mainly three possibilities: 
normal hierarchy (NH) with $m_1 \ll m_2 \ll m_3$,
inverted hierarchy (IH) with $m_3 \ll m_1 \simeq m_2$, or 
quasi-degenerate neutrinos (QD) with $m_0^2 = 
m_1^2 \simeq m_2^2 \simeq m_3^2 \gg \dms, |\dma|$. 
The latter requires that $m_{1,2,3} \gsim 0.10$ eV. 
For the QD case also one can still ask the question 
whether $m_1$ or $m_3$ is the lowest mass, i.e.~whether 
$\dma > 0$ or $\dma < 0$.  

In the present article we investigate in a 
model-independent way the possibility of having a sizeable 
value of $|U_{e3}|$ as a result of a perturbation of  
tri-bimaximal neutrino mixing. For concreteness, 
we will use the range of $|U_{e3}|$ in Eq.~(\ref{eq:bari}) 
in our analysis. In the light of expected deviations 
from TBM, it represents an 
interesting and testable benchmark scenario for various 
breaking mechanisms. 
In this respect, the problem of possible deviations of the neutrino mixing matrix 
from the TBM form has not been studied in detail (see e.g.~Ref.~\cite{ich} 
for some qualitative statements on the subject).  
Very specific perturbations to TBM in the framework 
of concrete models,  with the goal of allowing 
sizable non-zero $\theta_{13} \simeq 0.1$, have recently been 
discussed in Refs.\ \cite{king_new}. 
However, a detailed, quantitative and model-independent 
analysis, in particular  
in the light of the recent hints for a non-zero $U_{e3}$, 
has not been performed before and in our opinion is at the present 
stage both timely and useful. 

More specifically, in this paper we 
consider values of $|U_{e3}| \simeq 0.1$
suggested by Eq.~(\ref{eq:bari}) 
and try to obtain them by starting from TBM.  
The main challenge is to keep at the same time 
$\sin^2 \theta_{23}$, and especially $\sin^2 \theta_{12}$, 
close to their experimentally determined  
and thus close to the TBM predicted values. 
As any breaking mechanism introduces correlations between the
observables, we are able to make characteristic and testable 
predictions within each case. Interestingly, all predictions are 
connected with CP properties of the lepton sector.\\ 

The paper is organized as follows.
In Section \ref{sec:lep} we will start by deviating TBM with charged
lepton corrections and find that CP violation in neutrino oscillations
gets constrained to be almost maximal by the joint requirement of 
large $|U_{e3}|$ and small deviations from 
$\sin^2 \theta_{12} = \frac 13$. Atmospheric mixing deviates from maximal 
by order $\sin^2 \theta_{23} = \frac 12 + {\cal O}(|U_{e3}|^2)$. 
Section \ref{sec:RG} 
deals with quantum corrections to TBM and shows that only 
quasi-degenerate neutrinos in the Minimal Supersymmetric Standard
Model (MSSM) can give rise to sizable $|U_{e3}|$, while it is 
impossible to produce the required $|U_{e3}|$ if the effective 
theory is the standard model (SM).
Solar neutrino mixing is particularly affected by
renormalization effects, but the modification of 
$\theta_{12}$ can be suppressed by 
certain values of the Majorana CP violating phases. 
These values in turn 
influence the magnitude of the effective Majorana mass in 
neutrino-less double beta ($\betabeta$-)decay, 
leading to large cancellations. It is worth noting that 
the quantum corrections, within the context of the MSSM, 
make $\sin^2 \theta_{12}$ increase, 
whereas the $1\sigma$ range obtained 
from global fits lies below $\frac 13$. 
Atmospheric neutrino mixing 
deviates in general from maximal stronger than in the case of
charged lepton corrections, 
namely $\sin^2 \theta_{23} = \frac 12 + {\cal O}(|U_{e3}|)$. 
A similar but weaker correlation between $|U_{e3}|$ 
and the effective Majorana mass in $\betabeta$-decay  
is found when we explicitly perturb a neutrino mass 
matrix which without perturbations would lead to TBM. 
This possibility is analyzed in Section \ref{sec:expl}. 
We also find that in this case sizable corrections to 
$\sin^2 \theta_{23} = \frac 12$ of order $|U_{e3}|$ are expected. 
We finally summarize and conclude in  Section \ref{sec:concl}. 

%
\section{\label{sec:lep}Breaking Tri-bimaximal Mixing with Charged
Lepton Corrections}
%
%
The PMNS matrix is, in general, a product of two unitary matrices, 
\be
U = U_\ell^\dagger \, U_\nu\,,
\ee
%
where $U_\nu$ diagonalizes the neutrino mass matrix and $U_\ell$ 
is associated with the diagonalization of the charged lepton mass
matrix. Several authors have discussed charged lepton 
corrections to various neutrino mixing scenarios 
\cite{FPR,AK,otherslep,PR,HPR}. 
It has been shown \cite{FPR} that,
after eliminating the unphysical phases,
the matrix which diagonalizes the neutrino mass 
matrix can be written as: 
\be
U_\nu = P_\nu \, \tilde{U}_\nu \, Q_\nu \,,
\ee
where $\tilde{U}_\nu$ is a ``PDG-like'' mixing matrix, i.e., 
\be
\tilde{U}_\nu = \left(
\bad  
c_{12}^\nu \, c_{13}^\nu & s_{12}^\nu \, c_{13}^\nu & s_{13}^\nu 
\, e^{- i \xi}
\\[0.2cm] 
-s_{12}^\nu  \, c_{23}^\nu - c_{12}^\nu  \, s_{23}^\nu 
\, s_{13}^\nu  \, e^{i \xi} 
& c_{12}^\nu  \, c_{23}^\nu - s_{12}^\nu  \, s_{23}^\nu  
\, s_{13}^\nu  \, e^{i \xi} 
& s_{23}^\nu  \, c_{13}^\nu 
\\[0.2cm] 
s_{12}^\nu  \, s_{23}^\nu - c_{12}^\nu  \, c_{23}^\nu  
\, s_{13}^\nu  \, 
e^{i \xi} & 
- c_{12}^\nu  \, s_{23}^\nu - s_{12}^\nu  \, c_{23}^\nu  
\, s_{13}^\nu  \, e^{i \xi} 
& c_{23}^\nu  \, c_{13}^\nu 
\ea 
\right).
\ee
%
It contains three angles,
$\theta_{12}^{\nu}$, $\theta_{23}^{\nu}$, 
and $\theta_{13}^{\nu}$,
 and one phase, $\xi$. The diagonal matrices 
$P_\nu = {\rm diag}(1,e^{i \phi}, e^{i \omega})$ and 
$Q_\nu = {\rm diag}(1,e^{i \sigma}, e^{i \tau})$, in general,
cannot be neglected. Note, however, that $Q_\nu$ 
does not affect neutrino oscillation
observables \cite{BHP80,Lang86}. The unitary
matrix $U_\ell$ which diagonalizes the charged lepton 
mass matrix can be written as 
\bea
\label{eq:Ulep}
\tilde U_\ell 
= \left(
\bad  
c_{12}^\ell \, c_{13}^\ell & s_{12}^\ell \, c_{13}^\ell 
& s_{13}^\ell \, e^{- i \psi}
\\[0.2cm] 
-s_{12}^\ell \, c_{23}^\ell - 
c_{12}^\ell \, s_{23}^\ell \, s_{13}^\ell \, 
e^{i \psi} 
& c_{12}^\ell \, c_{23}^\ell - s_{12}^\ell \, s_{23}^\ell \, 
s_{13}^\ell 
\, e^{i \psi} 
& s_{23}^\ell \, c_{13}^\ell   
\\[0.2cm] 
s_{12}^\ell \, s_{23}^\ell - c_{12}^\ell \, c_{23}^\ell \, 
s_{13}^\ell 
\, e^{i \psi} 
& - c_{12}^\ell \, s_{23}^\ell - s_{12}^\ell \, c_{23}^\ell \, 
s_{13}^\ell \, 
e^{i \psi} 
& c_{23}^\ell \, c_{13}^\ell 
\ea 
\right) .
\eea
%
We have used in $U_{\nu, \, \ell}$ the obvious abbreviations 
$c_{ij}^{\ell, \nu} = \cos \theta_{ij}^{\ell, \nu}$ and 
$s_{ij}^{\ell, \nu} = \sin \theta_{ij}^{\ell, \nu}$. \\

Let us assume next 
that $U_\nu$ corresponds to TBM, i.e.,  
$\tilde{U}_\nu$ is given by $U_{\rm TBM}$ from Eq.~(\ref{eq:UTBM}). 
Assume further that the charged lepton corrections are ``CKM-like'',
i.e.~that
\be
\sin \theta_{12}^\ell = \lambda~,~~
\sin \theta_{23}^\ell = A \, \lambda^2~,~~
\sin \theta_{13}^\ell = B \, \lambda^3 \, ,
\ee
%
with $A, B$ real and of order one. We therefore have 
in mind here a GUT-like scenario, in which tri-bimaximal mixing from the 
neutrino sector (presumably owing its origin from a see-saw mechanism) 
is corrected by $U_\ell$, which via some quark-lepton
symmetry is related to the CKM mixing. A natural 
expectation for $\lambda$ is then that it is kindred to the sine of
the Cabibbo angle, $\lambda \simeq \sin \theta_C \simeq 
0.227$. In scenarios based on 
$SU(5)$ Grand Unification it often happens that a Clebsch-Gordan 
coefficient of $\frac 13$ occurs in between the charged lepton
and down quark diagonalization, in which case $\lambda \simeq 
\frac 13 \sin \theta_C \simeq 0.076$. 

In the case of CKM-like corrections it 
is straightforward to calculate from $U = U_\ell^\dagger \,
U_\nu$ the neutrino mixing observables 
$\sin^2 \theta_{12} = |U_{e2}|^2/(1 - |U_{e3}|^2)$, 
$\sin^2 \theta_{23} = |U_{\mu 2}|^2/(1 - |U_{e3}|^2)$ and 
$\sin \theta_{13} = |U_{e3}|$. Moreover, it is of interest to 
obtain the rephasing invariant  
\be
J_{\rm CP} = {\rm Im}\left\{ 
U_{e1}^\ast \, U_{\mu 3}^\ast \, U_{e 3} \,
U_{\mu 1} \right\} \, ,
\label{eq:JCP}
\ee
%
which controls the magnitude of CP violation in 
neutrino oscillations \cite{PKSP3nu88}, 
generated by the Dirac CP violating phase in 
the PMNS matrix. 
In the standard PDG-parametrization of the PMNS matrix we have 
$J_{\rm CP} = \frac 18 \, \sin 2 \theta_{12} \, \sin 2 \theta_{23} \,
\sin 2 \theta_{13} \, \cos \theta_{13} \, \sin \delta$. 
The result for the observables is 
\bea \label{eq:res_lep}
\sin^2 \theta_{12} \simeq \frac 13 
\left( 
1 - 2 \, \lambda \, \cos \phi + \frac 12 \, \lambda^2 
\right) ~,~~
|U_{e3}| \simeq \frac{ \D \lambda}{ \D \sqrt{2}} ~,\\  
\sin^2 \theta_{23} \simeq \frac 12 \left( 
1 - \left(\frac 12 - 2 \, A \, \cos (\omega - \phi) \right) \lambda^2
\right) ~,~~
J_{\rm CP} \simeq \frac 16 \, \lambda \, \sin \phi ~,
\eea
%
plus terms of order $\lambda^3$. 
The magnitude of $|U_{e3}|$ is in our analysis fixed by the range in 
Eq.~(\ref{eq:bari}). Therefore we can estimate the following 
interesting range for $\lambda$: $\lambda \simeq 0.18_{-0.07}^{+0.05}$. 

We can be more general, however, and refine this analysis. 
To this end, we consider the exact and lengthy
expression for $|U_{e3}|$ and use a random number generator to 
generate the values of $\lambda, A, B, \omega, \phi,
\psi$. We let $\lambda$ vary between 0 and 0.3, 
the phases between 0 and $2\pi$, and $A, B$ within 0.2 and 5. 
In order to have a hierarchy in $U_\ell$,
we take care that $\sin \theta_{12}^\ell$ is at least five
times as large as $\sin \theta_{23}^\ell$, which in turn is at least
five times as large as $\sin \theta_{13}^\ell$. We obtain then from the 
requirement of reproducing the $1\sigma$ ranges 
of the mixing angles given in 
Eqs.~(\ref{eq:data}, \ref{eq:bari}) the range 
\be \label{eq:lam_range}
0.104 \le \lambda \le 0.247\,.
\ee
%
This is the range for $\lambda$ we will use for the rest of this
Section. Interestingly, the sine of the Cabibbo angle is included in
this range, while one third of it is not. 
It turns out that $\sin^2 \theta_{12}$ can lie anywhere 
in its currently allowed range given in Eq.~(\ref{eq:data}). 
In contrast, as can be seen in Eq.~(\ref{eq:res_lep}), atmospheric 
mixing receives only small corrections of order $\lambda^2$, i.e., 
$\sin^2 \theta_{23} = \frac 12 + {\cal O}(|U_{e3}|^2)$. 
To be quantitative, we find  
\be \label{eq:lep_atm}
0.437 \le \sin^2 \theta_{23} \le 0.533 \, .
\ee
%
From the expressions for the mixing parameters 
given in Eq.~(\ref{eq:res_lep}), an interesting correlation 
appears \cite{PR}: 
a sizable value of $|U_{e3}|$, and therefore of $\lambda$, 
introduces a sizable contribution to $\sin^2 \theta_{12}$ of the same
order \footnote{Similar result holds in the case of
hierarchical $U_{\ell}$ and 
$U_\nu$ having bimaximal mixing form \cite{FPR}.}. To be more 
precise, we have: 
\be \label{eq:corr_lep}
\frac 13 - \sin^2 \theta_{12} \simeq \frac{2\sqrt{2}}{3} \, 
|U_{e3}| \, \cos \phi \, .
\ee
%
The observed value of $\frac 13 - \sin^2 \theta_{12} $ 
is at $1\sigma$ between $0.002$ and $0.039$. Thus, $\cos \phi$ should 
lie below 0.33, 0.26, or 0.53, if $|U_{e3}|^2 = 0.016$, 0.026, or 
0.006. The closer $\sin^2 \theta_{12}$ is to $\frac 13$, 
the smaller $\cos \phi$ is. Consequently, 
$|\sin \phi|$ is close to one and CP violation in neutrino 
oscillation is ``maximal'', in the sense that the invariant describing
it takes (as a function of $|U_{e3}|$) almost its maximal 
value \footnote{Solar neutrino mixing then receives correction by the 
``NLO'' term in Eq.~(\ref{eq:res_lep}): $\sin^2 \theta_{12} = \frac 13 
\left(1 + \frac 12 \, \lambda^2 \right)$.}. 

\begin{figure}[t]
\begin{center}
\parbox{3in}{
\epsfig{file=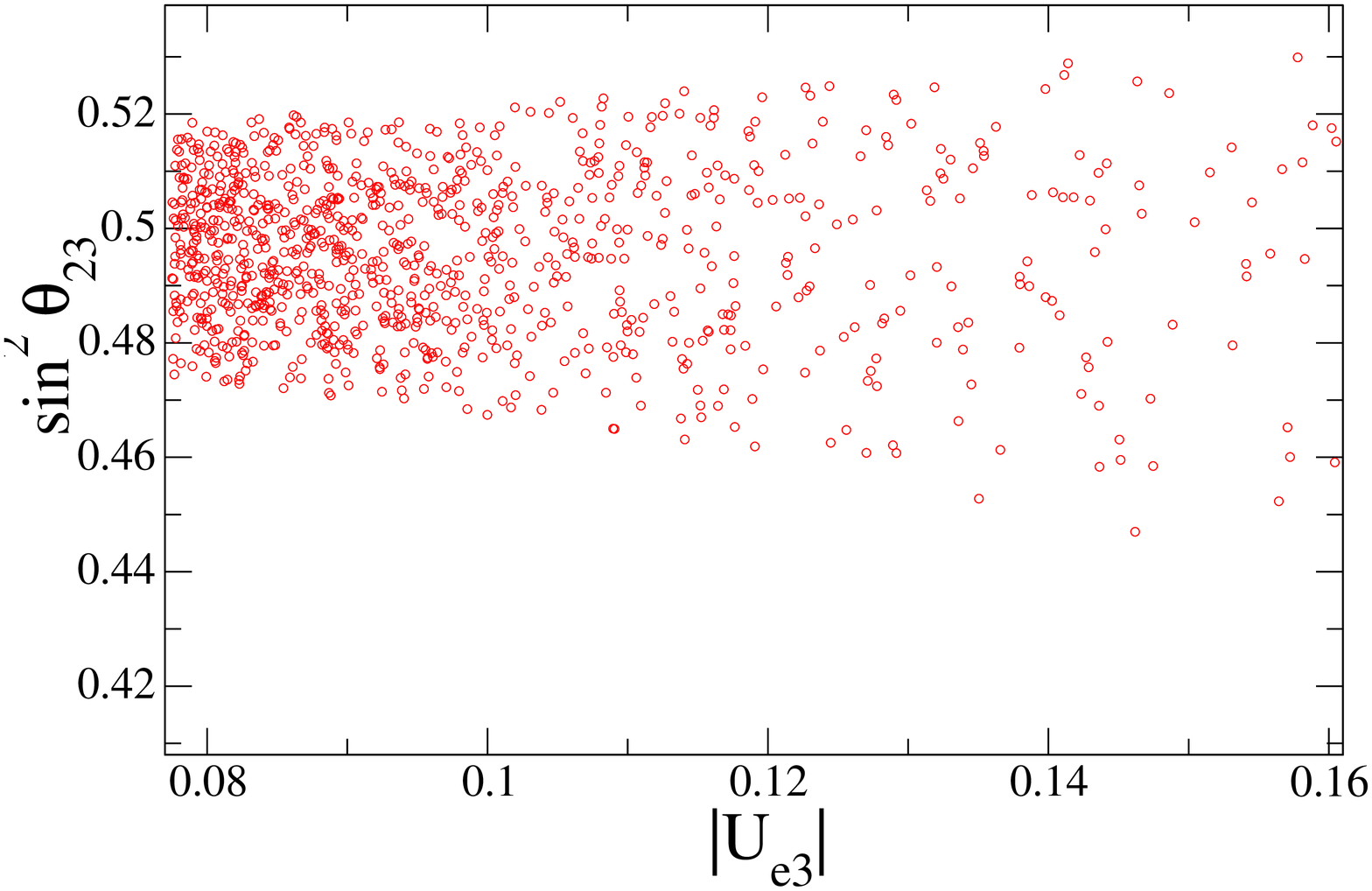,width=3in}
}
\parbox{3in}{
\epsfig{file=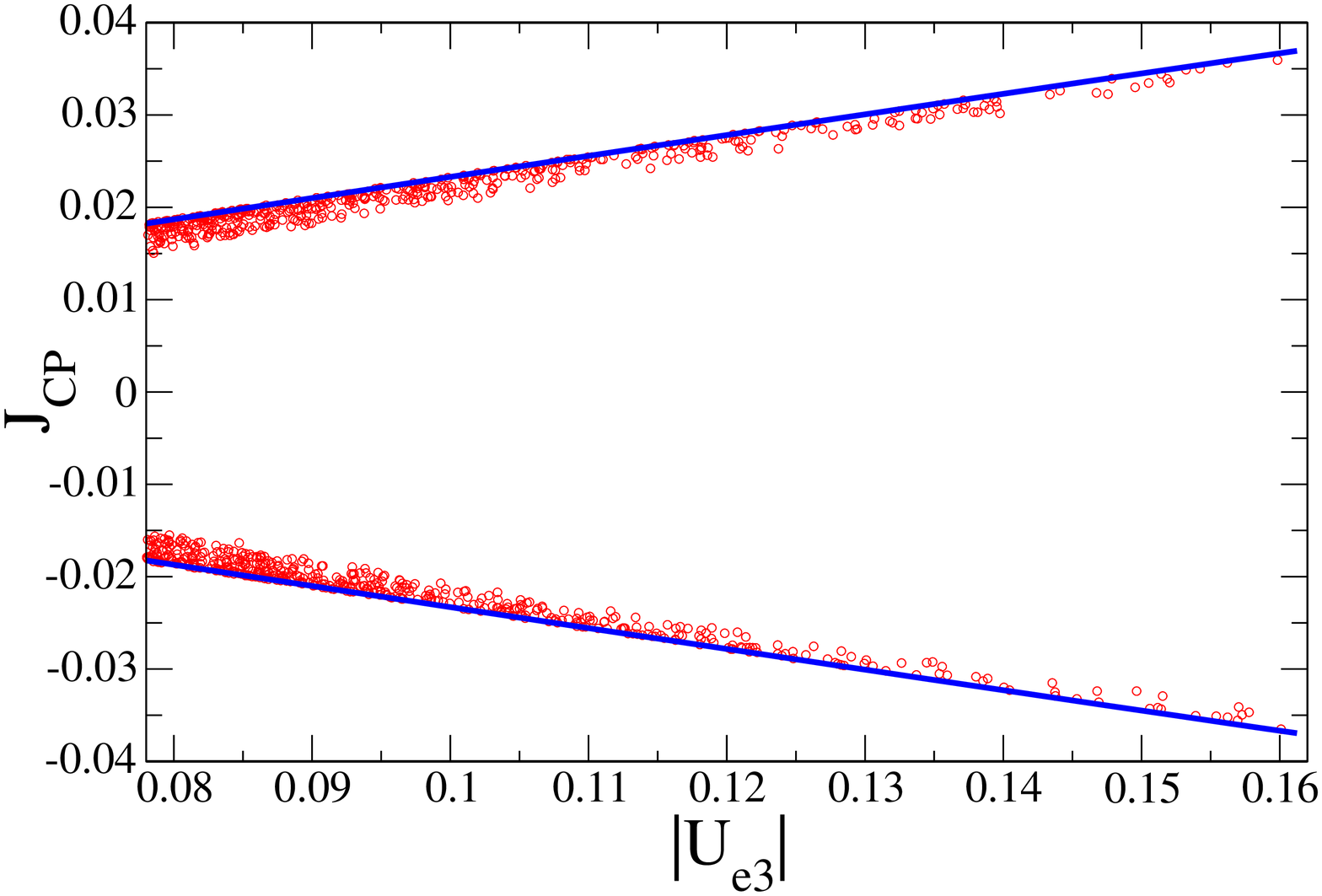,width=3.0in}
}
\caption{\label{fig:lep}Charged lepton corrections to tri-bimaximal
mixing. The left plot shows $|U_{e3}|$ against $\sin^2 \theta_{23}$ 
(the axes cover the whole $1\sigma$ range) 
while the right plot gives $|U_{e3}|$ against $J_{\rm CP}$. The blue
solid lines display the maximal 
value that $|J_{\rm CP}|$ can take. }
\end{center}
\end{figure}

 We illustrate the phenomenology of this 
framework in Fig.~\ref{fig:lep}. The values of the parameters in
$U_\ell$ are the same as the ones leading to 
Eqs.~(\ref{eq:lam_range}) and (\ref{eq:lep_atm}). It is easy to see
that $\sin^2 \theta_{23}$ can have values 
in a limited interval, and that 
CP violation is very close to maximal, 
i.e., $\delta = \phi~{\rm mod} \, \pi \simeq \pi/2$ or $3\pi/2$. 
The blue solid lines in Fig.~\ref{fig:lep} display the maximal 
value that $|J_{\rm CP}|$ can take. 
The sign of $\sin \delta$ cannot be predicted,
because the charged lepton corrections to the CP conserving 
quantities $\sin^2 \theta_{12}$ and $|U_{e3}|$ fix only $\cos \phi$, 
whereas CP violation depends necessarily on $\sin \phi$. Note that 
atmospheric neutrino mixing can be maximal.\\ 

Finally, we note an alternative second 
type of correction from the relation $U = U_\ell^\dagger \, 
U_\nu$, namely when $U_\ell^\dagger$ corresponds to TBM and 
$U_\nu$ is CKM-like \cite{HPR}. 
In this case, $|U_{e3}| \simeq \sin \theta_{23}^\nu /\sqrt{3}$. 
The parameter $|U_{e3}|$ is therefore governed by the 
23-element of $U_\nu$, which by the same arguments as given 
above for the first case, is expected to be quite small, namely of order 
$\lambda^2$. 
Even for the lowest considered value of $|U_{e3}|^2 = 0.006$ this 
scenario would require that $\sin \theta_{23}^\nu = 0.134$, 
a comparably large number, given the GUT-inspired paradigm of 
``small corrections'' in the relation $U_\ell^\dagger \, U_\nu$. 
We will therefore not discuss this possibility further, 
except for noting two things. First, sizable corrections 
of order $\lambda$ would arise, in general, for $\sin \theta_{12}$. 
Suppressing them by choosing a specific value of a CP violating 
phase is possible, but this phase is not 
related to CP violation in neutrino oscillations. 
Second, there would also be a very similar correlation to 
Eq.~(\ref{eq:corr_lep}), namely 
\be \label{eq:corr_lep2}
\sin^2 \theta_{23} - \frac 12  \simeq \sqrt{2} \, 
|U_{e3}| \, \cos \phi \, .
\ee
%
Note that in this case the atmospheric neutrino mixing angle is 
correlated with $|U_{e3}|$ and CP violation in neutrino 
oscillations. We refer to 
Ref.~\cite{HPR} for more details on this mixing scenario. 

%
\section{Breaking Tri-bimaximal Mixing with Quantum Corrections
\label{sec:RG}}
%
%

 Another straightforward breaking mechanism is the application of 
renormalization group (RG) corrections to 
TBM \cite{PR,DGR1,DGR2,RG_TBM}, 
which is essential to be considered if 
the tri-bimaximal scenario is assumed to have been generated 
at some high energy scale. 
In contrast to charged lepton corrections the results now depend on 
the neutrino mass values and their ordering. 
In general, in the usual PDG-parametrization of the mixing matrix,  
the corrections to the mixing angles can be expressed as
\cite{ASP,DGR1,DGR2}: 
\be
\theta_{ij}^\lambda \simeq \theta_{ij}^\Lambda + C 
\, k_{ij} \, \Delta_{\tau} + {\cal O}(\Delta_\tau^2) \, , 
\label{theta_ij}
\ee
%
where $\Lambda$ is the high scale at which TBM is implemented 
and $\lambda$ is the low energy scale at which 
measurements take place. We will indicate high scale values by a superscript 
$\Lambda$ in the following, and omit for simplicity the superscript 
$\lambda$, which would indicate low scale values.  
Hence we have 
$\theta_{12}^\Lambda = \sin^{-1}\sqrt{1/3}$, $\theta_{23}^\Lambda = \pi/4$ 
and $\theta_{13}^\Lambda = 0$. 
We consider the RG evolution of the neutrino masses and the mixing parameters 
in the effective theory and for definiteness assume the high 
scale to be $\Lambda = 10^{12}$ GeV. 
The low scale is taken to be $\lambda = 10^2$ GeV when the effective 
theory is the Standard Model (SM), while we take $\lambda=10^3$ GeV 
when the effective theory at low energy is the MSSM. 
The constant $C$ in Eq.~(\ref{theta_ij}) is given by $C = -3/2$ 
for the SM and $C = +1$ for the MSSM. 
The result in Eq.~(\ref{theta_ij}) is 
obtained in first order in the parameter 
%
\beqa
\Delta_\tau &\equiv& \left\{ 
\begin{array}{l} 
\frac{m_\tau^2}{8 \pi^2 \, v^2}\, (1 + \tan^2 \beta) 
\, \ln \frac \Lambda\lambda \simeq 1.4 \cdot 10^{-5}~(1 + \tan^2 \beta)    
\quad \quad {\rm (MSSM)}\, , \\
 \frac{m_\tau^2}{8 \pi^2 \, v^2}\, 
\, \ln \frac \Lambda\lambda \phantom{(1 + \tan^2 \beta)} \simeq  1.5 \cdot 10^{-5} 
\phantom{(1 + \tan^2 \beta)}
\quad \quad {\rm (SM)}\, , \end{array} \right.
\label{delta-tau}
\eeqa
with $\Delta_{e, \mu}$ having been neglected 
since $m_{e,\mu} \ll m_\tau$ and 
the vev of the Higgs is taken to be $v/\sqrt{2} = 174$ GeV.

\begin{table}[t]
\begin{center} 
\begin{tabular}{|c|c|c|c|} \hline 
Model & mass ordering & $\sin^2\theta_{12}$ & $\sin^2 \theta_{23} $  
\\ \hline \hline
\multirow{2}{*}{SM} & $\Delta m^2_{31} > 0 $ 
& $\searrow$ & $\searrow$\\ \cline{2-4} 
 & $\Delta m^2_{31} < 0 $ & $\searrow$ & $\nearrow$ \\ \hline 
\multirow{2}{*}{MSSM} & $\Delta m^2_{31} > 0 $ & $\nearrow$ 
& $\nearrow$ \\ \cline{2-4}  
 & $\Delta m^2_{31} < 0 $ & $\nearrow$ &  $\searrow$\\ \hline
\end{tabular}
\caption{\label{tab:RG}Direction of RG correction to the observables 
$\sin^2 \theta_{12}$ 
and $ \sin^2 \theta_{23}$ for the SM and the MSSM and both possible 
neutrino mass orderings.}
\end{center}
\end{table}
The dependence on the neutrino mass and mixing parameters is 
encoded in \cite{Antusch-RG,DGR1,DGR2}
\beqa 
k_{12} \nonumber 
&=& \frac{\sqrt{2}}{6}
\, \frac{\left| m_1 + m_2 \, 
e^{i \alpha_2}\right|^2}{\dmsq_{21}} ,\label{eq:k12} \\
k_{23} 
&=&-\left( \frac{1}{3} \, \frac{\left| m_2 + m_3 \, e^{i(\alpha_3 - \alpha_2)}
\right|^2}{\dmsq_{32}}  
+ \frac{1}{6} \, \frac{\left| m_1 + m_3 \, e^{i\alpha_3}\right|^2}
{\dmsq_{31}} \right) \; , \label{eq:kij} \\ 
k_{13} &=& -\frac{\sqrt{2}}{6}\left( 
\frac{\left| m_2 + m_3 \, e^{i(\delta + \alpha_3 - \alpha_2)}\right|^2}{\dmsq_{32}} -  
\frac{\left| m_1 + m_3 \, e^{i(\delta + \alpha_3)}\right|^2}{\dmsq_{31}} 
- \frac{4 \, m_3^2 \, \dmsq_{21}}{\dmsq_{31} \, \dmsq_{32}} \, 
\sin^2{\frac{\delta}{2}} \nonumber 
\right),
\label{eq:k13}
\eeqa
%
where we have used $\theta_{ij} = \theta_{ij}^\Lambda$, 
which is correct up to ${\cal O}(\Delta_\tau)$, 
and $\dmsq_{ij} \equiv m_i^2 - m_j^2$. 
The running of the masses  has been neglected in the above 
expressions for the $k_{ij}$. 
The masses are decreasing from high to low scale and run as 
\be
|m_i| = I_K \left(|m_i^\Lambda| + \mu_i \, \Delta_\tau\right) \, ,
\ee
%
where $I_K$ is a scalar factor that depends on the 
SU(2) and U(1) gauge coupling
constants and the Yukawa matrix in the up quark sector 
\cite{ellis-lola,chankowski,Antusch-RG} and $\mu_i$ are ${\cal{O}}(1)$ 
numbers.  
Thus, neglecting the running of masses \footnote{Note 
that the masses appear in both the denominator and numerator of the 
$k_{ij}$.} introduces an error
${\cal{O}}(\Delta_\tau)$  in $k_{ij}$ and hence 
${\cal{O}}(\Delta_\tau^2)$ in $\theta_{ij}$.  
One also observes that for 
$|\Delta_\tau| \gsim ( (m_2^\Lambda)^2- (m_1^\Lambda)^2)/(m_0^\Lambda)^2$,
the ${\cal O}(\Delta_\tau^2)$ terms dominate over the ${\cal O}
(\Delta_\tau)$ terms in the evolution of  $m_2^2-m_1^2$ 
\cite{Antusch-RG}. For such cases  
Eqs.~(\ref{eq:kij}) will no longer be cogent. 
Thus, for the validity of these equations, we require 
$(m_0^\Lambda)^2  \, \Delta_\tau \lsim (m_2^\Lambda)^2 - (m_1^\Lambda)^2 $, 
which may not be satisfied if
$(m_2^\Lambda)^2 - (m_1^\Lambda)^2$ is indeed very small. 
We will therefore 
use the full running equations for the mass matrix itself
for the plots and numerical values to be presented.  
Analytical estimates 
are made with the expressions of the $k_{ij}$ and, as we show, 
these estimates  can explain the numerical results with a sufficient
degree of correctness. 

There is a subtle issue involved when we consider $k_{13}$ in 
Eq.~(\ref{eq:kij}). 
As is seen, $k_{13}$ depends on the Dirac CP phase $\delta$ which 
is unphysical for the case of $\theta_{13} = 0$ 
at the high scale $\Lambda$. 
However as discussed in \cite{Antusch-RG,theta13_zero}, 
the value of $\delta$ at this point depends on the values of the masses 
and the Majorana phases 
and RG evolution takes care of that automatically.

For
analytical estimates, it is convenient to consider the shift of the
mixing angles $\theta_{ij}$ from their initial values.
From the above
expressions for the $k_{ij}$, and in the limit of 
$|k_{ij} \, \Delta_\tau| \ll 1$, one obtains the following 
expressions for the observables: 
\be \label{eq:obs}
\left|\sin \theta_{13}\right| \simeq 
\left| C \, k_{1 3} \, \Delta_\tau \right|~ , ~
\sin^2 \theta_{23} \simeq \frac 12 - C \, k_{23} \, \Delta_\tau~,~
\sin^2 \theta_{12} -\frac{1}{3} \simeq 
\frac{2 \sqrt{2}}{3} \, C \, k_{12} \, \Delta_\tau ~.
\ee
%
In the spirit of our analysis we require 
(see Eqs.~(\ref{eq:data}, \ref{eq:bari})) that 
$|C \, k_{13} \, \Delta_\tau| = 0.077 - 0.161$, while 
$-C \, k_{12} \, \Delta_\tau = 2.8 \cdot 10^{-3} - 4.2 \cdot 10^{-2}$. 
Note that for the $1\sigma$ range we are taking, 
$C \, k_{12} \, \Delta_\tau$ (and therefore $C$)
is supposed to be negative.  
Hence, within the MSSM the required deviation from TBM 
cannot be realized. 
Therefore we use the 3$\sigma$ ranges for $\sin^2\theta_{12}$ 
and $\sin^2\theta_{23}$ for the purposes of illustration. 
If indeed the trend of $\sin^2 \theta_{12} < \frac 13$ continues then 
it will not be possible to account for 
a high scale value of $\sin^2 \theta_{12} = \frac 13$
 solely by RG effects 
within the MSSM \footnote{Note that for the same reason 
any initial value of $\sin^2 \theta_{12} > \frac 13$ 
(including bimaximal mixing) at 
high scale is excluded unless of course highly model-dependent see-saw 
threshold effects \cite{treh,theta13_zero} are taken into account.}.    

From Eqs.~(\ref{theta_ij}) and (\ref{eq:kij})  
it is evident (and well-known) that whether 
the angles $\theta_{ij}$ will decrease or increase during 
evolution will depend on the effective theory (SM or MSSM, 
through the factor $C$) 
and also on the sign of $\dmsq_{31} (\simeq \dmsq_{32})$ for $\theta_{23}$. 
Table \ref{tab:RG} summarizes the direction of the correction for 
the observables $\sin^2 \theta_{12}$ and $\sin^2 \theta_{23}$.  
Since $k_{12}$ is always positive,  
$\theta_{12}$ at the low scale is always
larger (smaller) than that at high scale for the MSSM (SM). 
The size of the RG corrections will depend on the values 
of the Majorana phases, the neutrino masses and, in 
case of the MSSM, $\tan \beta$.

One can at the outset make some interesting observations from 
Eqs.~(\ref{eq:kij}).  
It is easy to see that only quasi-degenerate neutrinos will be able 
to lead to values of $|U_{e3}|$ around 0.1. Note also that in this case 
the running of solar neutrino mixing is in general enhanced by a factor 
$|\dma|/\dms$ with respect to the running of the other mixing angles. 
We will see that suppressing the running of $\sin^2 \theta_{12}$ 
and hence of the factor $k_{12}$ by suitable values of the Majorana 
phase $\alpha_2$ around $\pi$ has interesting 
phenomenological consequences in  
$\betabeta$-decay.
We further can expect that the deviation from 
maximal $\theta_{23}$ is of the same order than the deviation from zero 
$|U_{e3}|$. In the following we will quantify these 
statements.

We have performed a detailed analysis, 
by numerically solving the RG running equations for the 
effective neutrino mass matrix and then  
diagonalizing it to extract the masses, mixing angles and phases 
at low scale.  
At the high scale $\Lambda$, the angles $\theta_{ij}^\Lambda$ are fixed 
by the requirement of the TBM scenario, while 
the masses and the CP phases are chosen randomly so that after 
the RG evolution at low scale the parameters are 
consistent with the chosen ranges of the current experimental data. 
We have used the following ranges for the high scale values of 
the mass-squared differences: $(\dms)^\Lambda = 10^{-6} - 10^{-3}$ eV$^2$ and  
$|\dma|^\Lambda =  1.5 \times 10^{-3} - 10^{-2}$ eV$^2$, while the 
phases are varied over the full range of $0 - 2 \pi$. 
\\ 

\begin{figure}[t]
\begin{center}
\epsfig{file=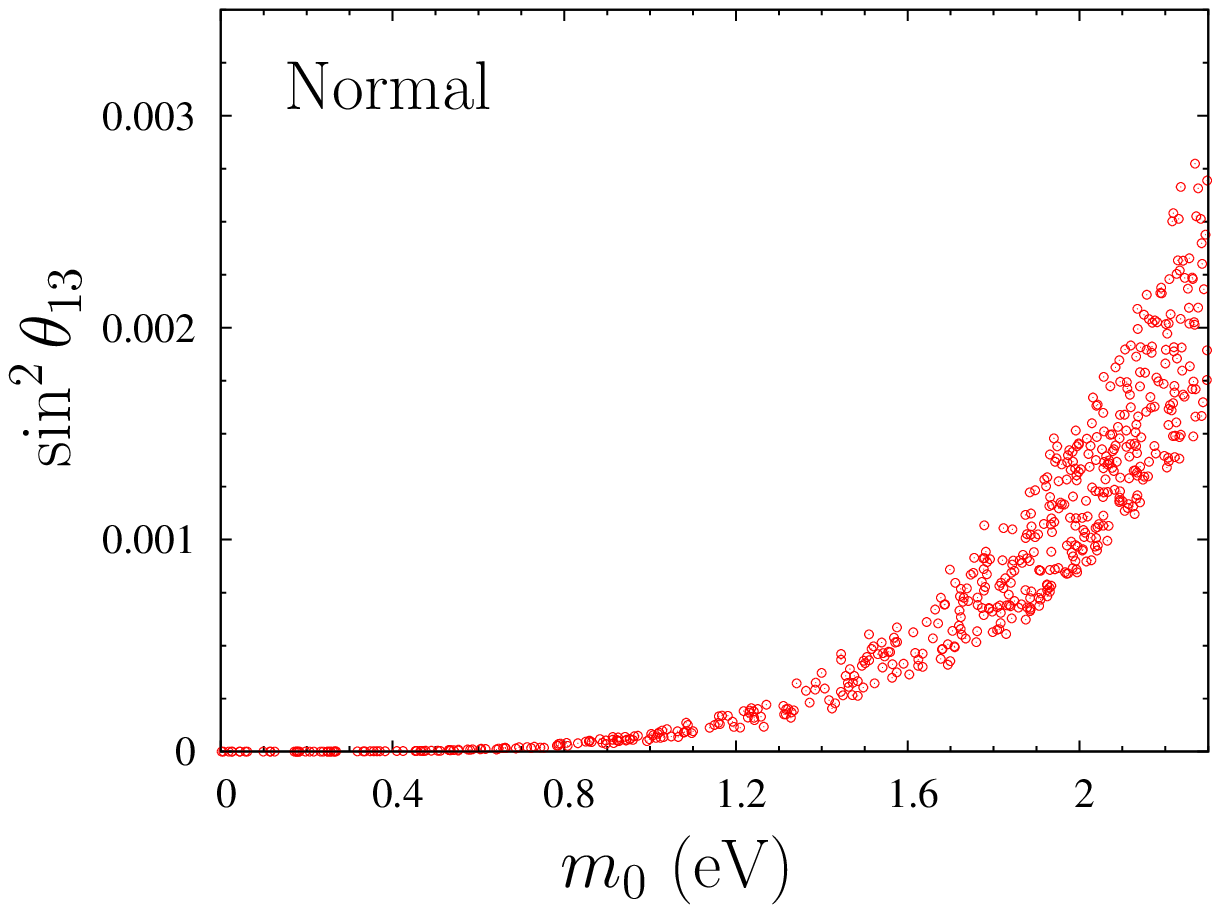,width=7.68cm,height=6cm}
\epsfig{file=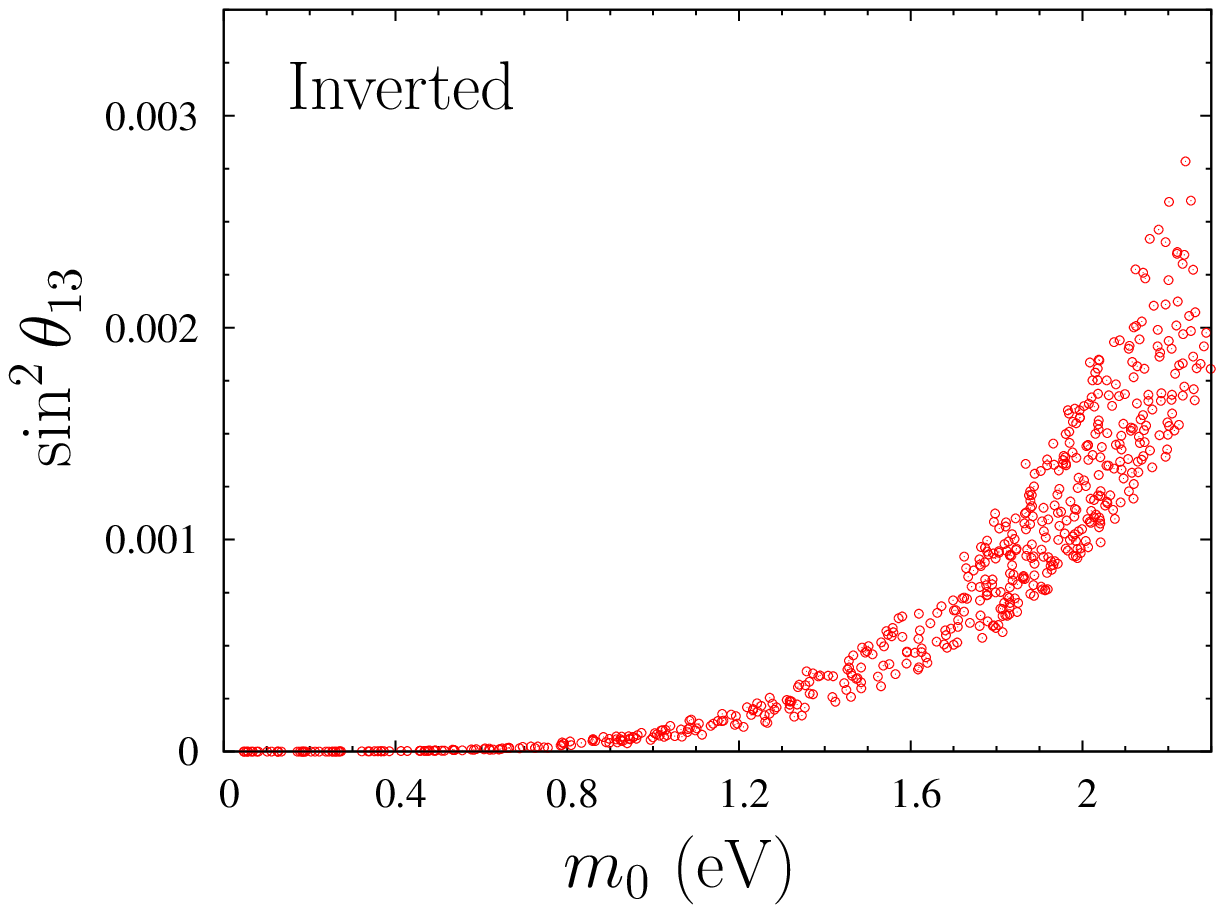,width=7.38cm,height=6cm}
\caption{\label{fig:SM}
The running of $|U_{e3}|^2 = \sin^2\theta_{13}$ 
in SM for both the normal (left panel) 
and the inverted (right panel) mass orderings. 
The high scale values of mixing angles are kept fixed at
TBM values while the masses and phases are varied randomly such 
that after RG evolution the parameter values are within current 
experimental ranges.  
}
\end{center}
\end{figure}

Starting with the SM, Fig.~\ref{fig:SM} shows the allowed region in the 
$m_0$ -- $\sin^2\theta_{13}$ plane 
at the low scale $\lambda$, after performing the RG evolution, for 
both the normal (left panel) and inverted (right panel) mass orderings. 
Recall that $m_0^2 \gg |\dma|$ 
is the common neutrino mass scale for quasi-degenerate 
neutrinos. 
As can be seen, to generate values of $|U_{e3}|$ within the range of interest, 
neutrino masses should exceed the direct limit of 2.3 eV 
from tritium decay \cite{mainz}, and hence also the more stringent but  
model-dependent limits from cosmology.  
We conclude that a high scale value of 
$\theta_{13} = 0$ is incompatible with the indicated range of $|U_{e3}|$. 
The dependence of this statement on 
the initial values of $\theta_{12}^\Lambda$ and $\theta_{23}^\Lambda$ 
is moderate and hence this statement is valid in general.\\

We will focus on the MSSM in what follows. As already stated 
above, we require the $3\sigma$ ranges of the oscillation parameters to be 
satisfied, because, strictly speaking, the MSSM cannot reproduce the 
$1\sigma$ range, due to its prediction of $\sin^2 \theta_{12} \ge \frac 13$.  
\begin{figure}[t]
\begin{center}
\epsfig{file= 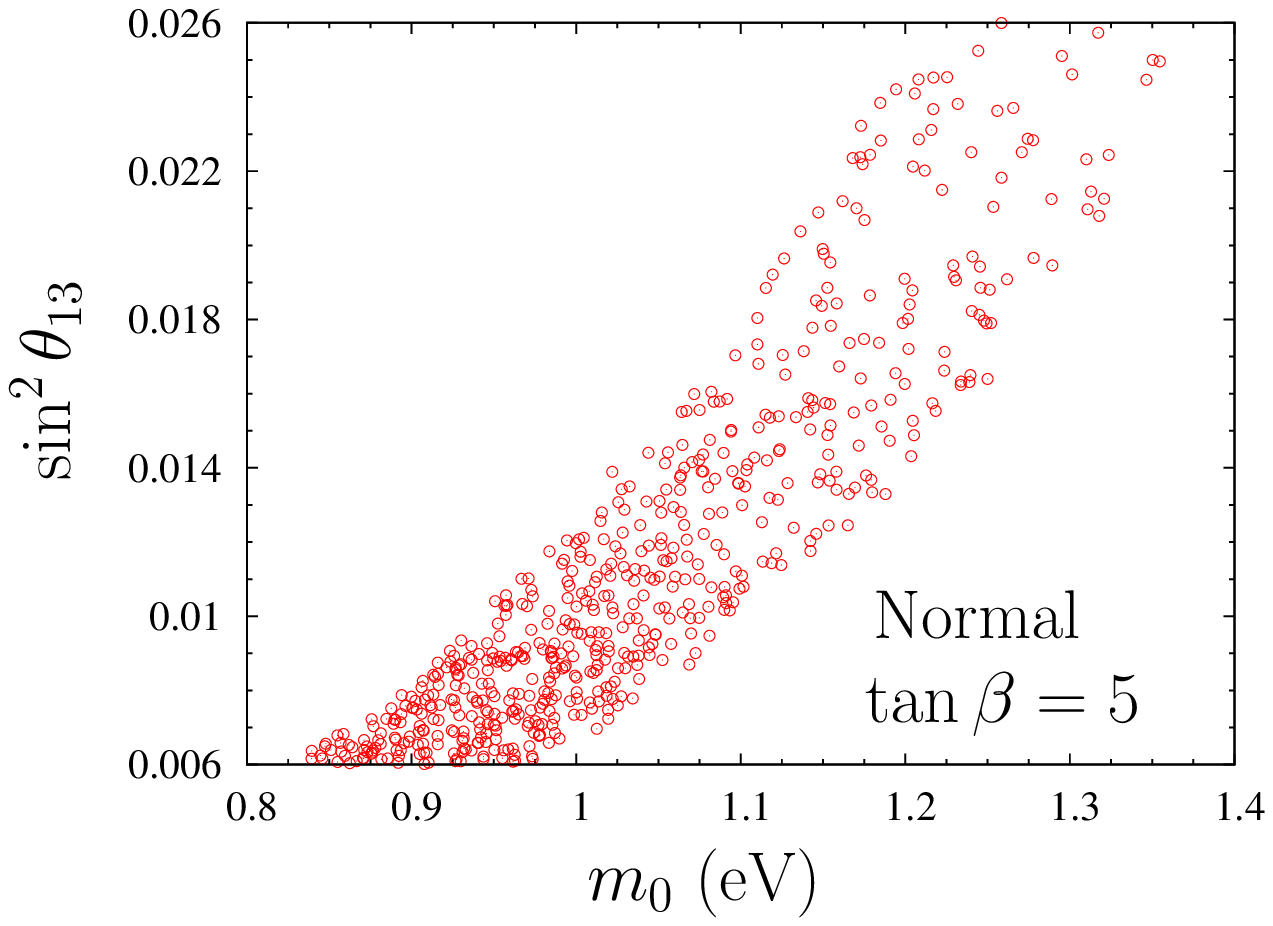,width=7.68cm,height=6cm}
\epsfig{file= 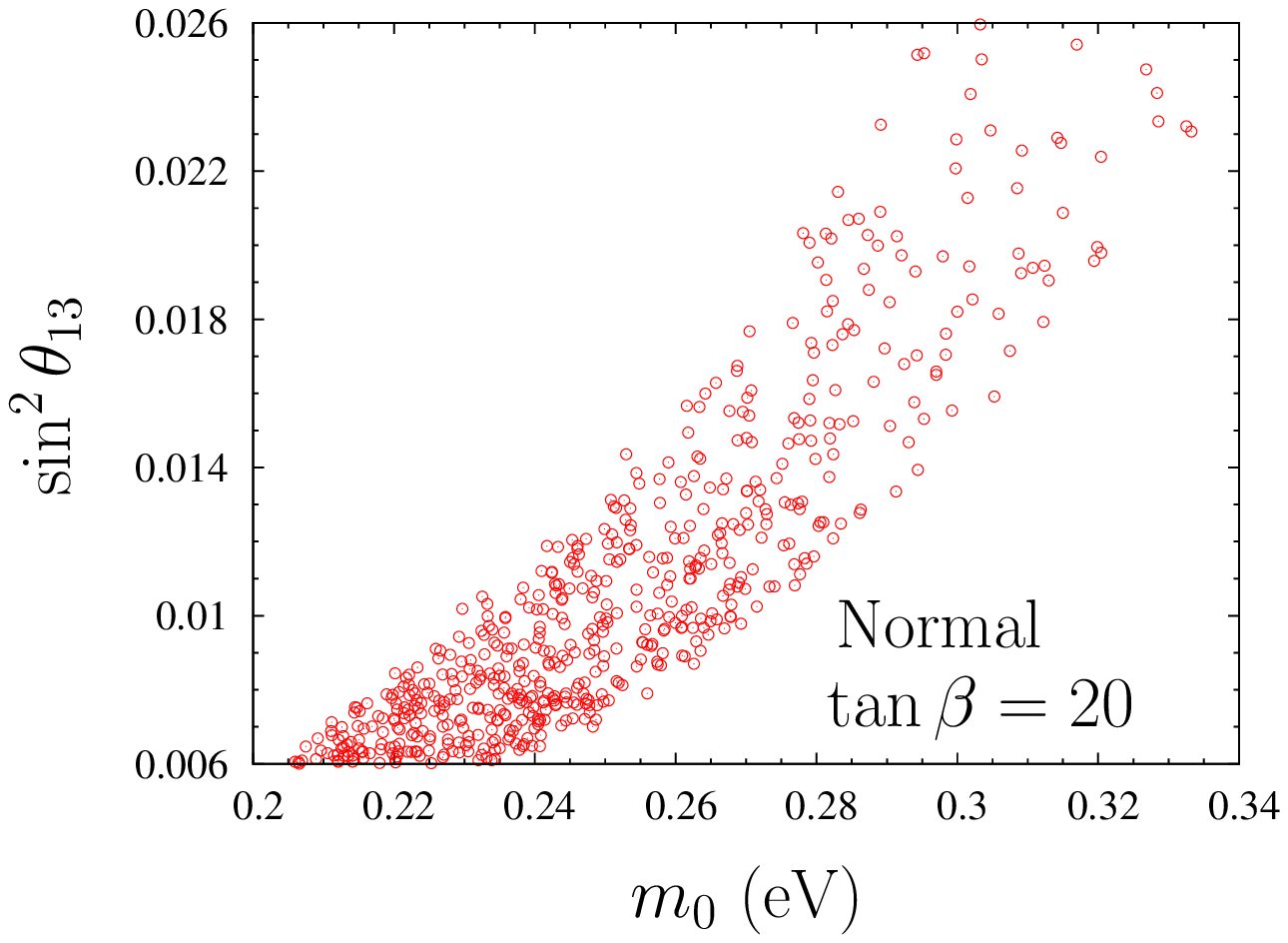,width=7.68cm,height=6cm}
\caption{\label{fig:MSSM-RG-m0-sinsqth13}
Scatter plots showing the running of $\sin^2\theta_{13}$ with 
$m_0$, for MSSM with normal mass ordering and $\tan\beta=5,20$. 
The high scale mixing angles are fixed at TBM and the masses and 
phases are varied randomly such that after RG evolution the 
parameter values are within current experimental ranges. For 
a given $\tan\beta$, the allowed regions are the same as above for 
the inverted mass ordering.} 
\end{center}
\end{figure}
%

Fig.~\ref{fig:MSSM-RG-m0-sinsqth13} shows the allowed region in 
the $m_0$ -- $\sin^2 \theta_{13}$ plane, 
when the effective theory is the MSSM, for 
$\tan\beta = 5,20$ and the normal mass ordering. The left panel shows that 
$\sin^2\theta_{13}$ lies in the required range when 
0.8 eV $\ls m_0 \ls$ 1.4 eV for $\tan\beta = 5$, while the 
allowed mass range becomes 0.2 eV $\ls m_0 \ls$ 0.34 eV 
for $\tan\beta=20$, as can be seen from the right panel. 
Thus, the relevant range of $m_0 \tan\beta$ is given by 
(see below for analytical estimates) 
$4.1 \ls (m_0/{\rm eV}) \tan\beta \ls 6.9$. 
Hence the allowed mass ranges depend strongly 
on $\tan\beta$ and for higher values of $\tan\beta$, lower values 
of $m_0$ are sufficient to produce the 
required running of $\theta_{13}$. 
It has been checked that for a fixed $\tan\beta$ value, 
there is no significant dependence on the mass ordering, other than 
the direction of the correction to $\theta_{23}$. 
From the allowed mass ranges obtained in 
Fig.~\ref{fig:MSSM-RG-m0-sinsqth13} 
it is seen that to have $\sin^2\theta_{13}$ in the 1$\sigma$ range under 
consideration, we 
need the neutrinos to be quasi-degenerate even for the 
MSSM with $\tan\beta=20$.  
\begin{figure}[t]
\begin{center}
\epsfig{file= 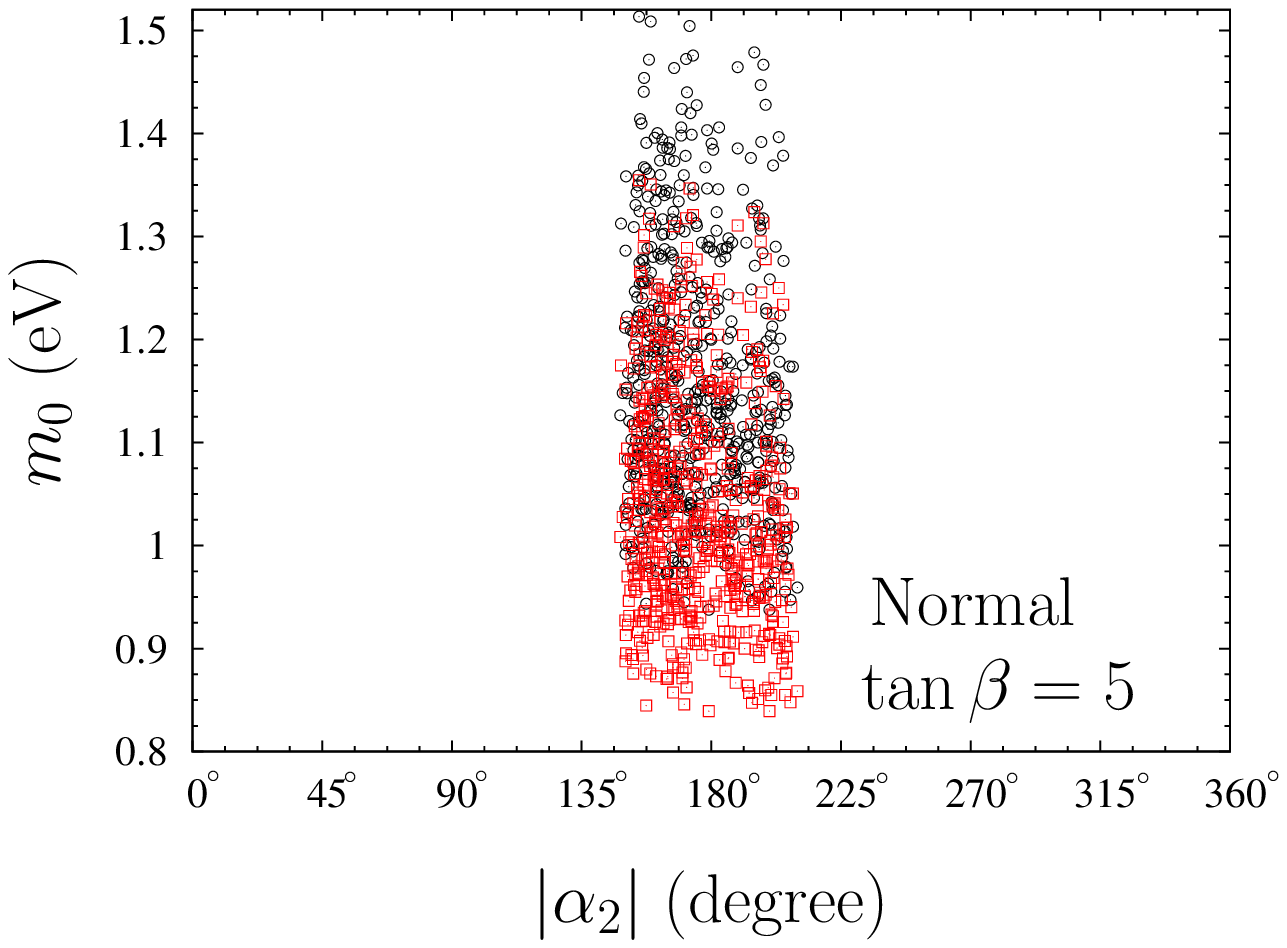,width=7.68cm,height=6cm}
\epsfig{file= 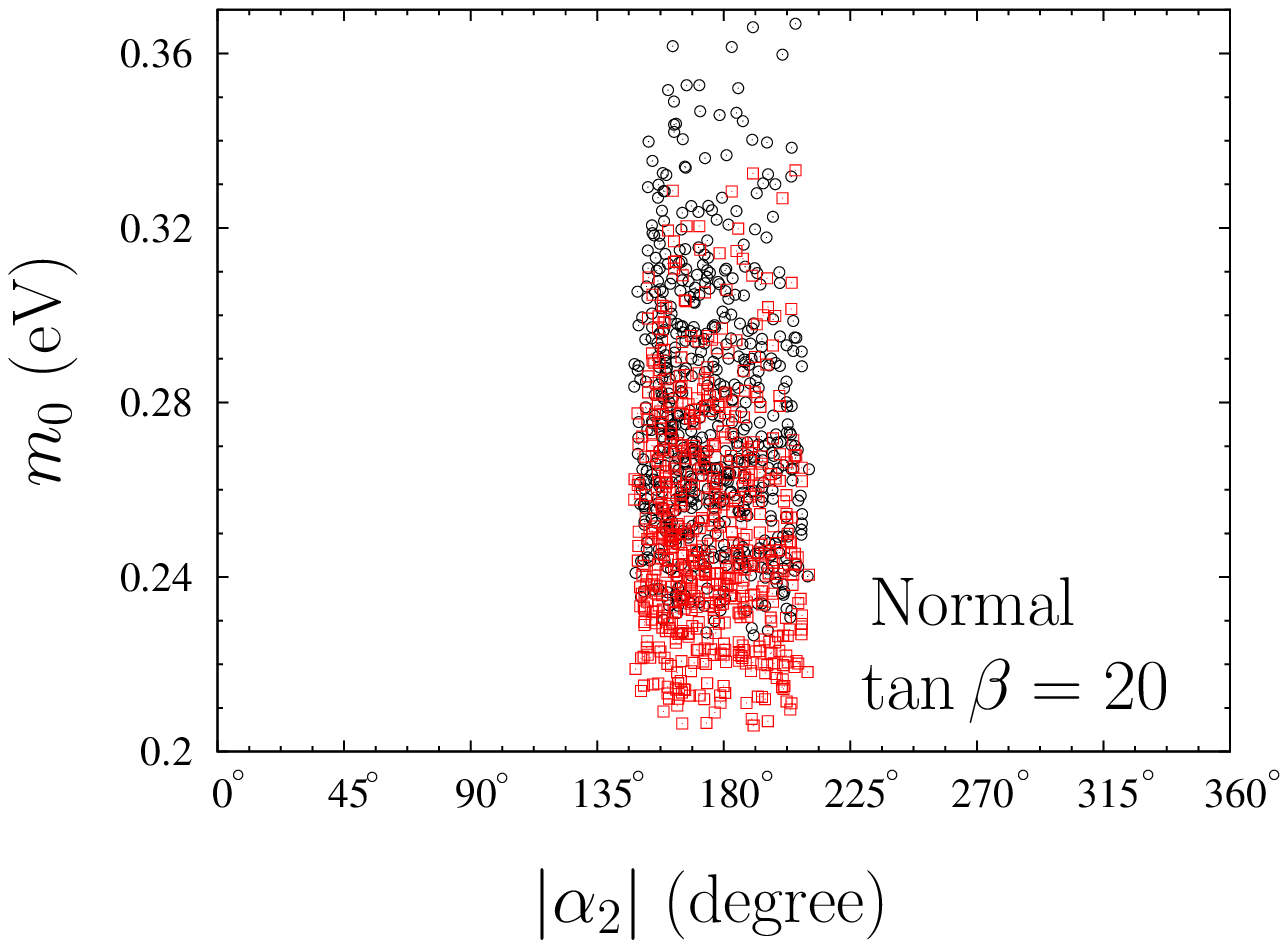,width=7.68cm,height=6cm}
\caption{Scatter plots in the $m_0$ -- $|\alpha_2|$ plane 
for MSSM ($\tan\beta=5,20$) and normal mass ordering, 
both for high (black circles) and low (red squares) energy scales. 
The high scale mixing angles are fixed at the TBM values 
and the masses and phases are varied randomly at the high scale
so that the low energy parameters are consistent with the 
current experimental data. The data for the inverted 
mass ordering shows the same variation. 
\label{fig:MSSM-RG-m0-alpha2}} 
\end{center}
\end{figure}
%
Fig.~\ref{fig:MSSM-RG-m0-alpha2} shows scatter plots of 
the allowed region of the neutrino mass scale $m_0$ 
and the Majorana phase $\alpha_2$, which is particularly 
important for the running of $\theta_{12}$ \cite{Antusch-RG}
(see also \cite{PShinYasu05}). 
We compare the allowed regions 
at high and low scale for a normal mass ordering and 
$\tan \beta = 5, 20$. The scattered plots 
obtained for the inverted mass ordering show the same characteristics. 
We see that $|\alpha_2|$ is restricted in a narrow region around 
$|\alpha_2| = \pi$ for all cases.

In order to explain the plots analytically we consider Eqs. (\ref{eq:k12}) 
in the QD regime $m_0^2 \gg \dma$ to obtain: 
\be \D \label{eq_t12QD}
\left(\sin^2 \theta_{12} - \frac 13 \right)_{\rm QD} \simeq
\frac 49 \, C \, \Delta_\tau \, (1 + \cos \alpha_2) \,
\frac{m_0^2}{\dms} \, ,
\ee
\be \label{eq:t13QD} 
|\sin\theta_{13}|_{\rm QD} \simeq \frac{\sqrt{2}}{3} \, C \, \Delta_\tau \,
\frac{m_0^2}{\dma} \,
\left| (1+R) \, \cos (\delta + \alpha_3 - \alpha_2) - \cos (\delta +\alpha_3) + 
R \, \cos \delta \right| \; , 
\ee
%
where $R = \dms/\dma$.  
From Eq.~(\ref{eq_t12QD}) one can understand that  
the low energy constraint on $\sin^2\theta_{12}$ from the current 
experimental data restricts $|\alpha_2|$ to remain close to $\pi$, 
as shown in Fig.~\ref{fig:MSSM-RG-m0-alpha2},  
making $\left(1+\cos \alpha_2\right)$ small so that  
there is less running of $\theta_{12}$ even with large neutrino masses. 
The plots in Fig.~\ref{fig:MSSM-RG-m0-alpha2} 
further show that $\alpha_2^\Lambda$ is also close to $\pi$ and that  
$\alpha_2$ stays close to $\pi$ in the course of its RG evolution. This 
can be estimated from the fact that   
the running of $\alpha_2$ can be expressed as 
$\alpha_2^\lambda \simeq \alpha_2^\Lambda + a_2 \, \Delta_\tau$ 
\cite{Antusch-RG,DGR2} with
$  a_2 \simeq - 2/(3 \, \dms) \, m_1^\Lambda \, m_2^\Lambda \,  
\sin \alpha_2^\Lambda$. 
%
%
\begin{figure}[t]
\begin{center}
\epsfig{file= 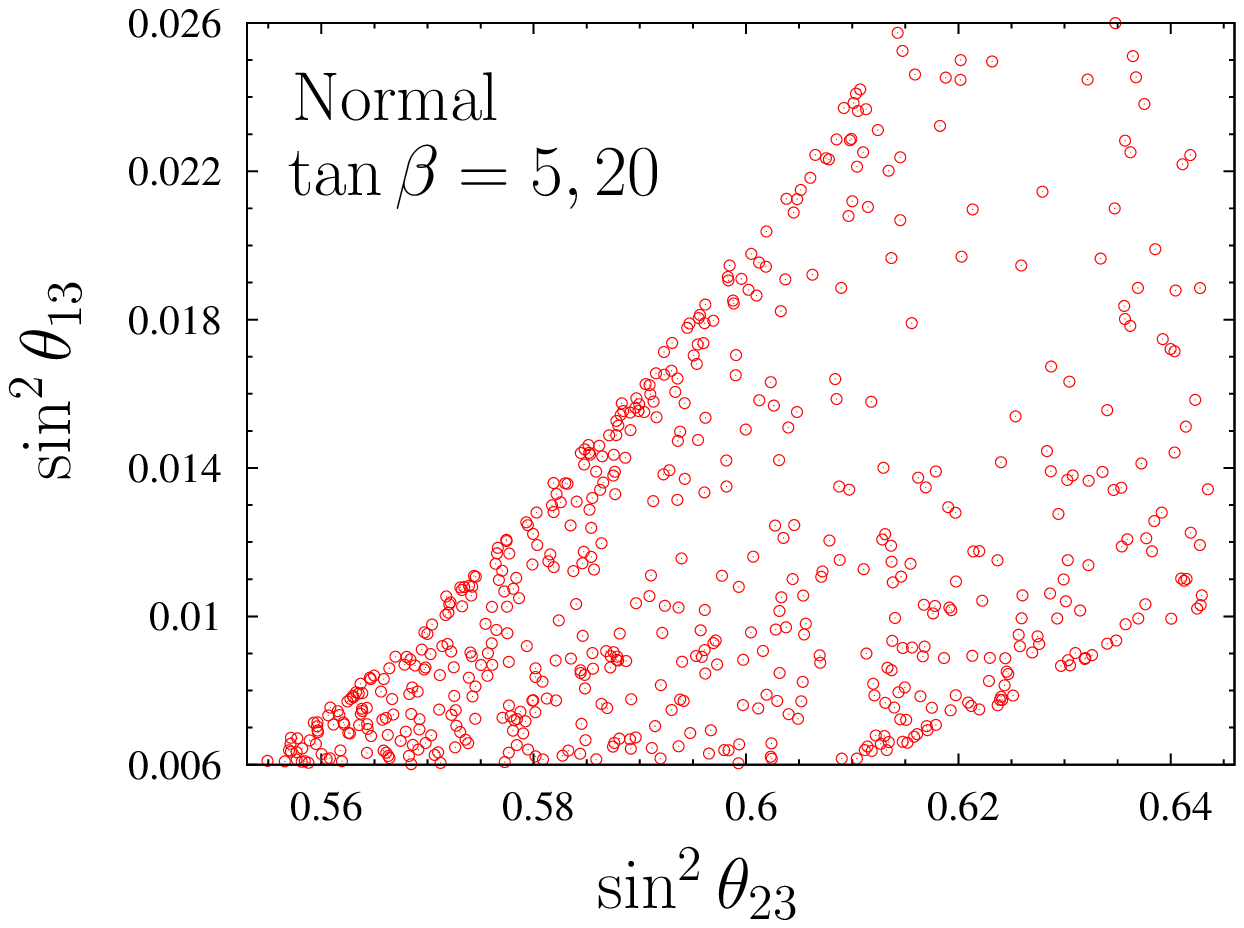,width=7.68cm,height=6cm}
\epsfig{file= 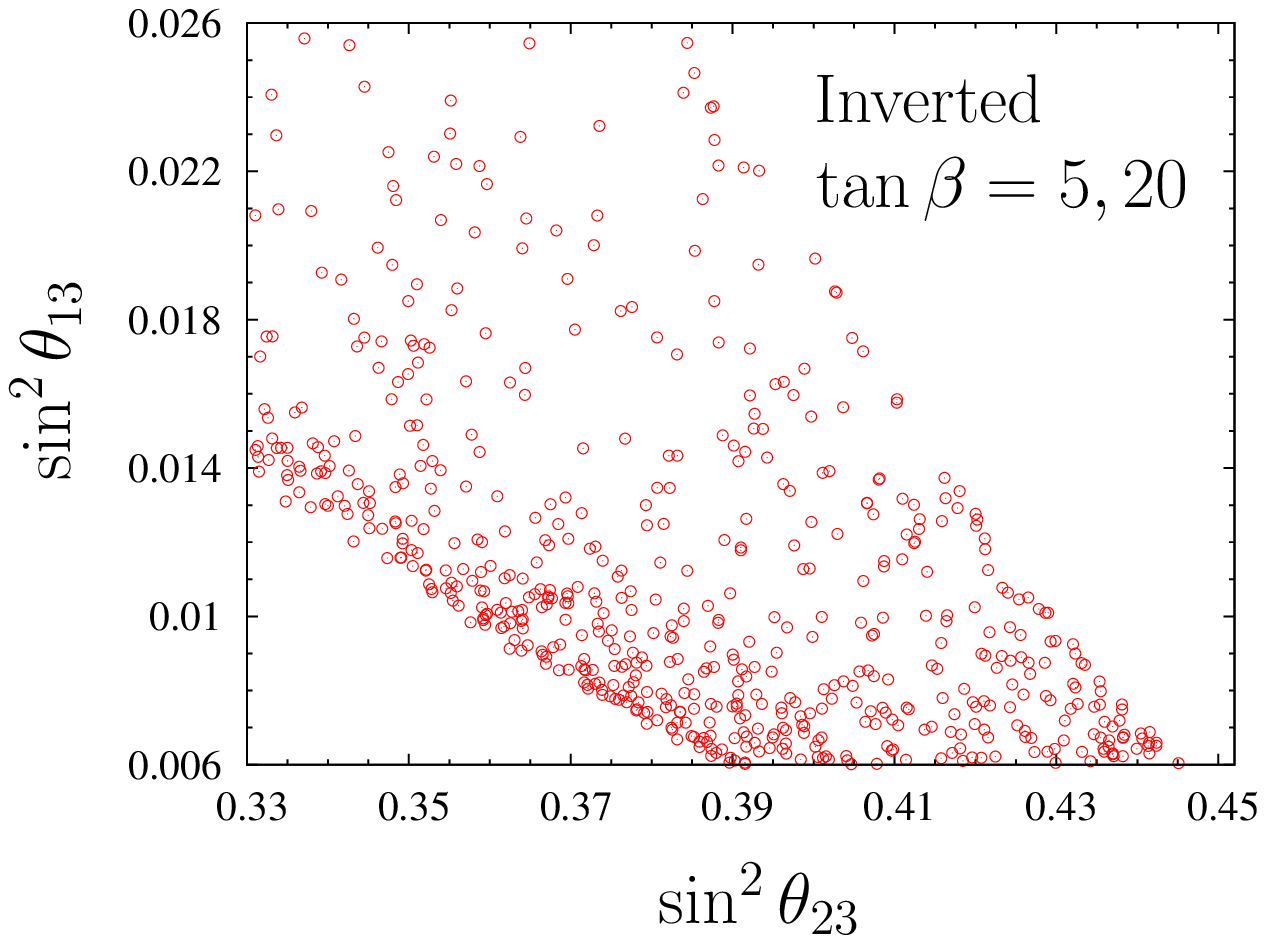 ,width=7.68cm,height=6cm} 
\caption{Scatter plots showing the correlation between  
$\sin^2\theta_{13}$  and $\sin^2\theta_{23}$ for   
MSSM with $\tan\beta=5,20$. The left panel shows the case 
with normal mass 
ordering, while the right panel is for the inverted mass ordering.
The high scale mixing angles are fixed at TBM and the masses and 
phases are varied randomly such that after RG evolution the 
parameter values are within current experimental ranges.
\label{fig:MSSM-RG-sinsqth23-sinsqth13}} 
\end{center}
\end{figure}
From Eq.~(\ref{eq_t12QD})  we note that the maximum running for 
$\theta_{12}$ is obtained for $\alpha_2=0$.
In absence of any lower bound on $\theta_{13}$ this value was still
allowed \cite{DGR2}.
However if we put $\alpha_2=0$ in Eq.~(\ref{eq:t13QD}) then the
running of $\theta_{13}$ is suppressed by the factor 
$|R| = \dms/|\dma|$. 
Thus, the requirement of large running of $\theta_{13}$
disfavors $\alpha_2=0$ and further strengthens
the bound in the $\alpha_2$ -- $m_0$ plane.\\

In the limit of $\alpha_2=\pi$ and quasi-degenerate neutrinos,
the maximum value of $|\sin\theta_{13}|$ that can be achieved starting 
from $\theta_{13}^\Lambda = 0$ can be estimated from 
Eq.~(\ref{eq:t13QD}) as
\be 
|\sin\theta_{13}|_{\rm QD} \leq \frac{2\sqrt{2}}{3} \, C \, 
\Delta_\tau \,
\frac{m_0^2}{\dma} \, (1 + R) \, ,
\label{sinsqth13-QD-a2pi}
\ee
with $\alpha_3 = 0$, $\delta = \pm \pi$ or 
$\alpha_3 = \pm \pi$, $\delta = 0$. 
Thus, from Eq.~(\ref{sinsqth13-QD-a2pi}) one can estimate 
that $\theta_{13} \ge 0.077$ 
requires $m_0 \gs$ 2.66 eV for the SM and $m_0 \gs$ 0.72 (0.18) eV 
for $\tan\beta = 5 \, (20)$ with the MSSM. The estimates are in good  
agreement with the allowed mass ranges obtained from 
Fig.~\ref{fig:SM} and Fig.~\ref{fig:MSSM-RG-m0-sinsqth13}, 
respectively.

Fig.~\ref{fig:MSSM-RG-sinsqth23-sinsqth13} shows the 
correlation between the low scale 
values of $\sin^2\theta_{13}$ and $\sin^2\theta_{23}$. 
For normal ordering $\sin^2\theta_{23} > \frac 12$, 
whereas for inverted ordering 
$\sin^2 \theta_{23} < \frac 12$. For normal ordering
$\theta_{13}$ and $\theta_{23}$ are correlated, i.e.,  
a higher value of $\theta_{13}$ requires a higher value of
$\theta_{23}$. 
\begin{figure}[t]
\begin{center}
\epsfig{file=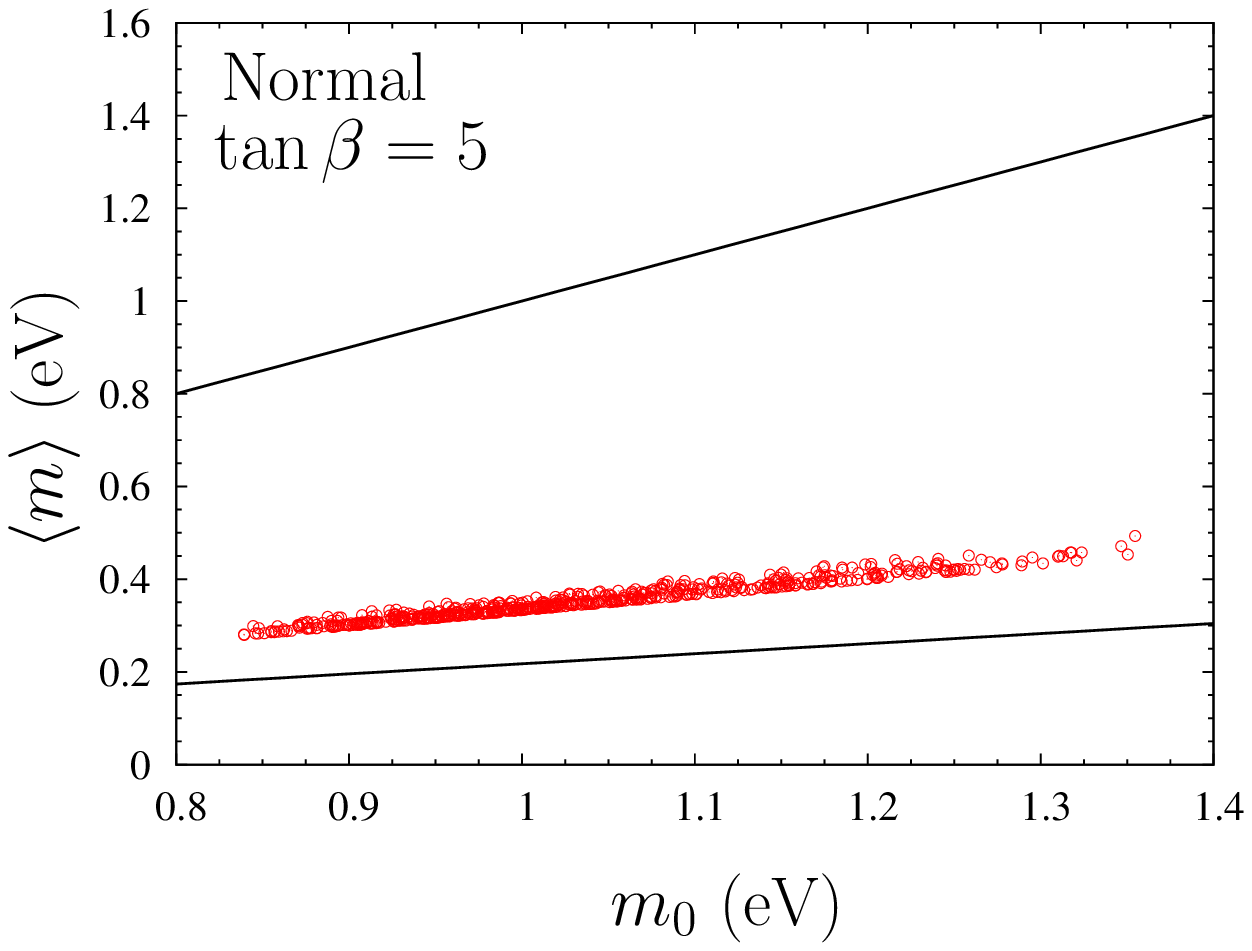,width=7.68cm,height=6cm}
\epsfig{file=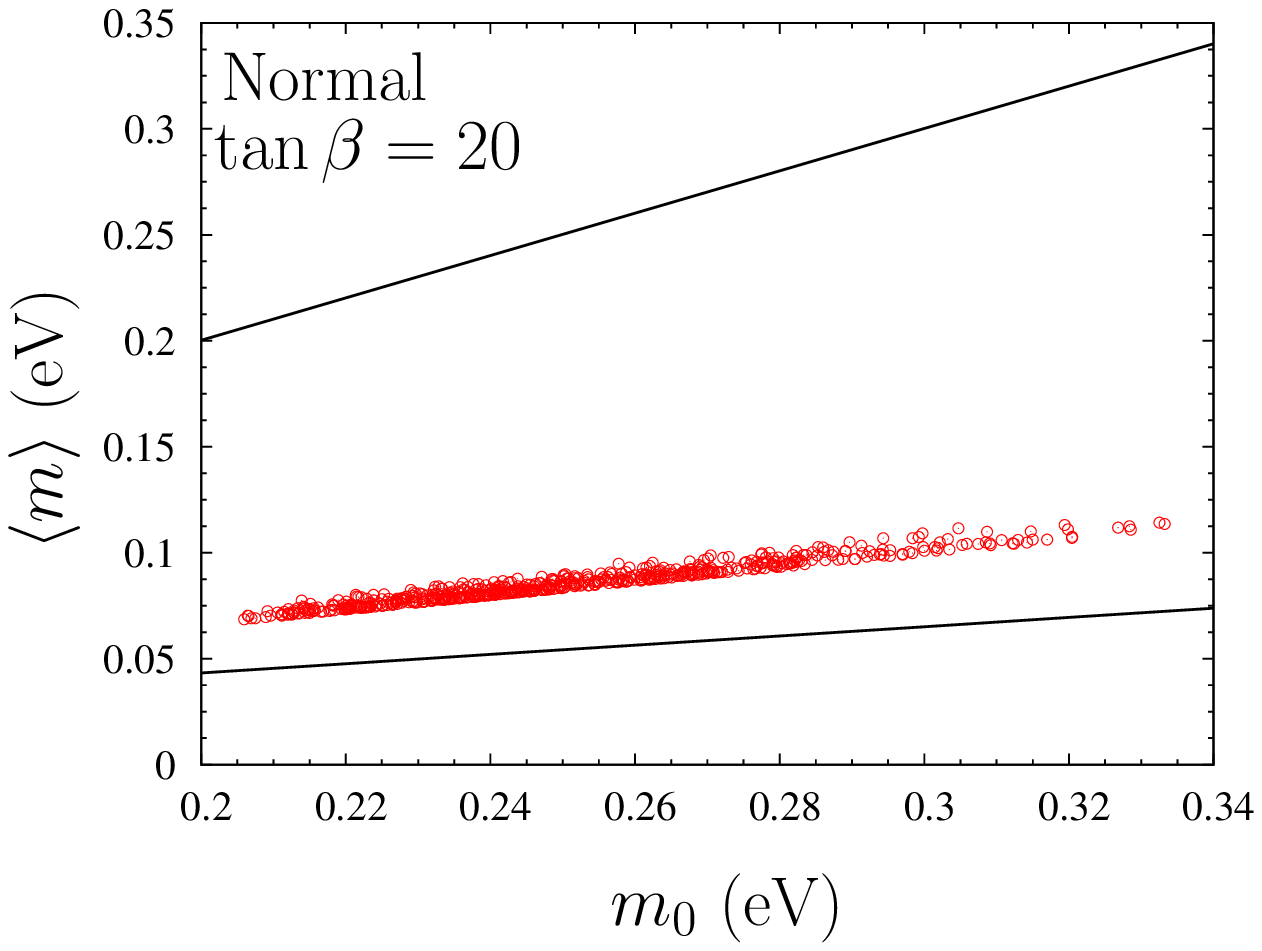,width=7.68cm,height=6cm}
\caption{\label{fig:MSSM-RG-m0-mneu}
Scatter plot for the effective neutrino mass $\meff$ that contributes
to neutrino-less double beta decay as a function of $m_0$, in MSSM with 
$\tan \beta = 5,20$ and normal mass ordering. 
The solid (black) lines indicate the maximum and minimum possible 
values of $\meff$ for given $m_0$, obtained by varying the 
oscillation parameters in their current $3\sigma$ range and the 
phases between 0 to $2 \pi$. The cases with inverted 
ordering show identical characteristics.}
\end{center}
\end{figure}
For the inverted ordering the predicted 
values of the two angles are anti-correlated. 
The plots obtained with $\tan\beta=20$ are identical to those  
shown in Fig.~\ref{fig:MSSM-RG-sinsqth23-sinsqth13} 
for $\tan\beta=5$, when the mass ordering is the same.
For a different $\tan\beta$ the value of $m_0$ adjusts itself 
to comply with the low energy cuts on the parameters  and the 
allowed points in the 
$\sin^2\theta_{23} - \sin^2\theta_{13}$ plane remain same. 

We note here that maximal atmospheric neutrino mixing is not
possible. To be more quantitative, we find that 
\be
\baz 
0.55 \le \sin^2 \theta_{23} \le 0.64 & \mbox{ for } 
\Delta m_{31}^2 > 0 \, ,\\
0.33 \le \sin^2 \theta_{23} \le 0.45 & \mbox{ for } \Delta m_{31}^2 < 0 \, ,
\ea
\ee
independent on the value of $\tan \beta$. 

In Fig.~\ref{fig:MSSM-RG-m0-mneu} we plot the effective Majorana mass 
\be \label{eq:meff}
\meff = \cos^2 \theta_{13} \left| 
m_1 \, \cos^2 \theta_{12} + m_2 \, \sin^2 \theta_{12} 
\, e^{i \alpha_2} + m_3 \, \tan^2 \theta_{13} \, e^{i (\alpha_3+2 \delta)} 
\right| ,
\ee
%
which governs the rate of 
$\betabeta$-decay at low energy. 
The scatter points show the values of $\meff$ allowed by the low 
energy neutrino oscillation data after RG analysis. 
The solid (black) lines indicate the maximum and minimum possible 
values of $\meff$ at low scale for a given $m_0$, obtained by varying the 
oscillation parameters in their current $3\sigma$ range and the 
phases between 0 to $2 \pi$. 
The plots show that the effective mass obtained after RG analysis 
lies close to its minimum allowed range. 
As can also be seen from Fig.\ \ref{fig:MSSM-RG-m0-mneu}, for 
$\tan \beta = 5$, \meff~takes values between 
0.26 and 0.50 eV, to be compared with the general upper and lower 
limits of 0.2 eV  and 1.4 eV. If $\tan \beta = 20$, then 
$0.07~{\rm eV}\ls \meff \ls 0.11~{\rm eV}$, 
while in general the effective Majorana mass could be 
in between 0.05 eV and 0.34 eV.  
As can be estimated from Eq.~(\ref{eq:meff}), 
the maximum value that ${\meff}$ 
can achieve for quasi-degenerate 
neutrinos is when $\alpha_2 = \alpha_3 + 2 \delta = 0$, and  
is given by \cite {BPP1} $m_0$, 
whereas the minimal value is 
obtained when $\alpha_2 = \alpha_3 + 2 \delta = \pi$:  
\be
{\meff}_{\rm QD}^{\rm min} \simeq 
m_0 \, \left(\cos 2\theta_{12} 
- 2 \, |U_{e3}|^2/(1 + \tan^2 \theta_{12}) \right) \, .  
\ee
%
As we have seen, 
RG evolution combined with low energy constraints imply 
QD neutrinos with $\alpha_2$  close to $\pi$. In this limit   
\beqa
{\meff}_{\rm QD}^{\alpha_2=\pi} \simeq m_0 \, \cos^2 \theta_{13} \left| 
\cos2\theta_{12} + \tan^2 \theta_{13} \, e^{i (\alpha_3+2 \delta)} 
\right| 
\label{mee-QD-RG-1}\; ,
\eeqa
%
and since $\theta_{13}$ is small at all energy scales, expanding in 
powers of $|U_{e3}| = \sin\theta_{13}$ we can write \cite{BPP1}
\beqa
{\meff}_{\rm QD}^{\alpha_2=\pi} \simeq m_0  
\left( \cos2\theta_{12} + {\cal O}(|U_{e3}|^2) \right)
\label{mee-QD-RG} \; .
\eeqa
%
Thus,  
RG evolution constrains 
$\meff$ towards the  minimum allowed value, which is confirmed by the 
figure. \\

We can give very simple forms of the 
neutrino mass matrix in the flavor basis satisfying the above
constraints. In general the mass matrix generating TBM reads 
\be \label{eq:mnutbm}
(m_\nu)_{\rm TBM} = 
U_{\rm TBM}^\ast \, P^\ast \, m_\nu^{\rm diag}\,P^\dagger\, 
U_{\rm TBM}^\dagger = 
\left(
\bad 
A & B & B \\[0.2cm]
\cdot & \frac{1}{2} (A + B + D) & \frac{1}{2} (A + B - D)\\[0.2cm]
\cdot & \cdot & \frac{1}{2} (A + B + D)
\ea 
\right)\,.
\ee
%
The parameters $A,B,D$ are in general complex and functions 
of the neutrino masses and Majorana phases:
\be \D 
A = \frac 13 \left(2 \, m_1 + m_2 \, e^{-i\alpha_2} \right) \,,~~
B = \frac 13 \left(m_2 \, e^{-i\alpha_2} - m_1 \right) \,,~~
\D D = m_3 \, e^{-i\alpha_3} \,.
\ee
%
Note that the sum of the elements in each row, and in each column,
equals $A + 2 B = m_2 \,e^{-i\alpha_2}$. Now to estimate the 
texture of $m_\nu$ at high scale let us insert $m_{1,2,3} = m_0$, 
TBM and $\alpha_2 = \pi$. It follows 
\[ 
\frac{3}{m_0} \, m_\nu \simeq   
\left( 
\bad
1 & -2 & -2 \\
\cdot & \frac 12 \left(-1 + 3 \, e^{-i \alpha_3} \right) & 
-\frac 12 \left(1 + 3 \, e^{-i \alpha_3} \right) \\ 
\cdot & \cdot & \frac 12 \left(-1 + 3 \, e^{-i \alpha_3} \right)
\ea 
\right) \rightarrow 
\left\{ 
\baz 
\left( 
\bad
1 & -2 & -2 \\
\cdot & 1 & -2 \\
\cdot & \cdot & 1
\ea \right) & \mbox{ for } \alpha_{3} = 0 \, ,\\
\left( 
\bad
1 & -2 & -2 \\
\cdot & -2 & 1 \\
\cdot & \cdot & -2
\ea \right) & \mbox{ for } \alpha_{3} = \pi \, ,
\ea \right. 
\]
%
where we have set two specific values of $\alpha_3$. Corrections to these 
expressions are of order $\sqrt{\dms}/m_0$, $\sqrt{\dma}/m_0$ 
and hence small for QD neutrinos. 

\begin{figure}[t]
\begin{center}
\parbox{3in}{
\epsfig{file=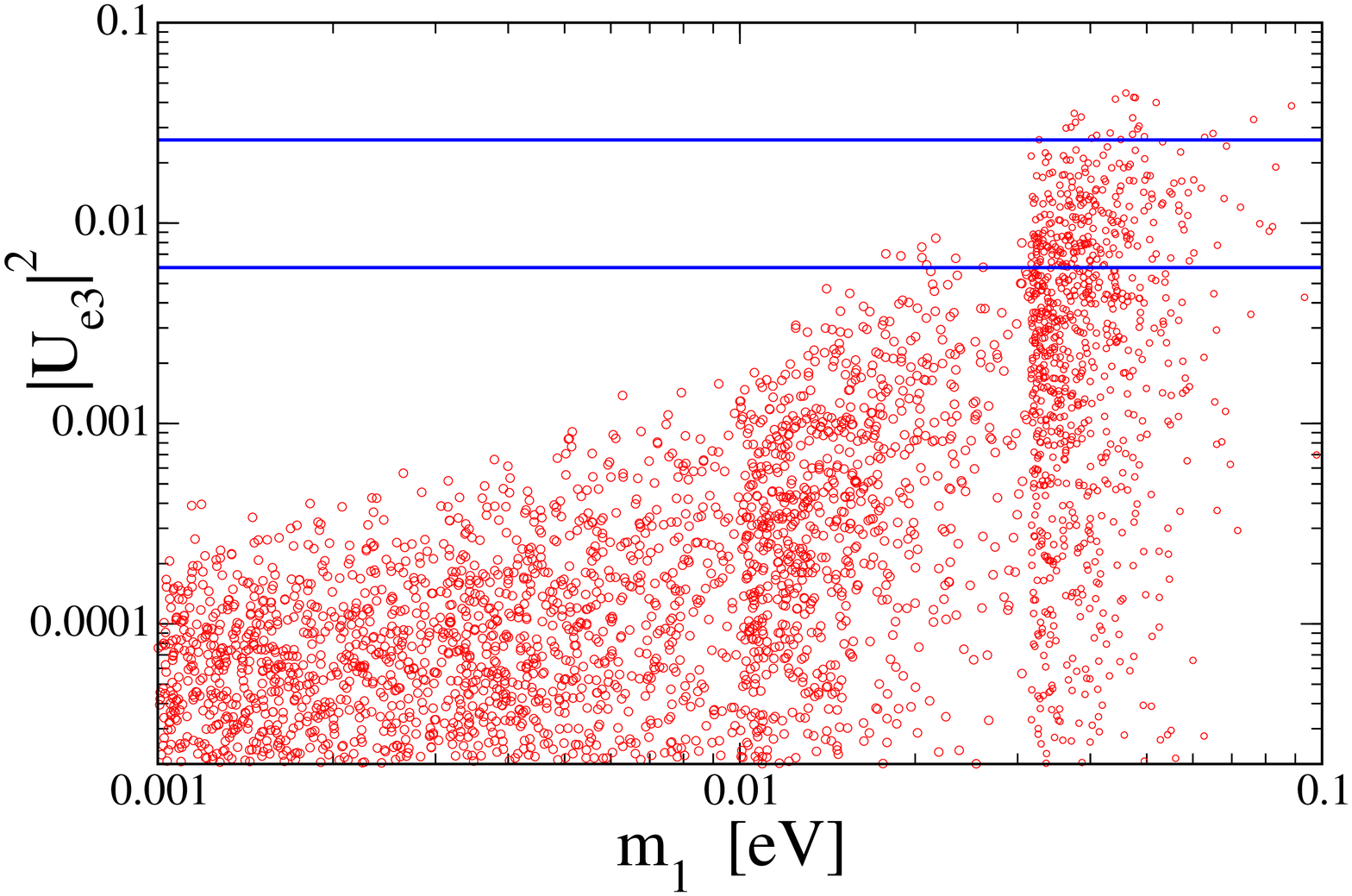,width=3in}
}
\parbox{3in}{
\epsfig{file=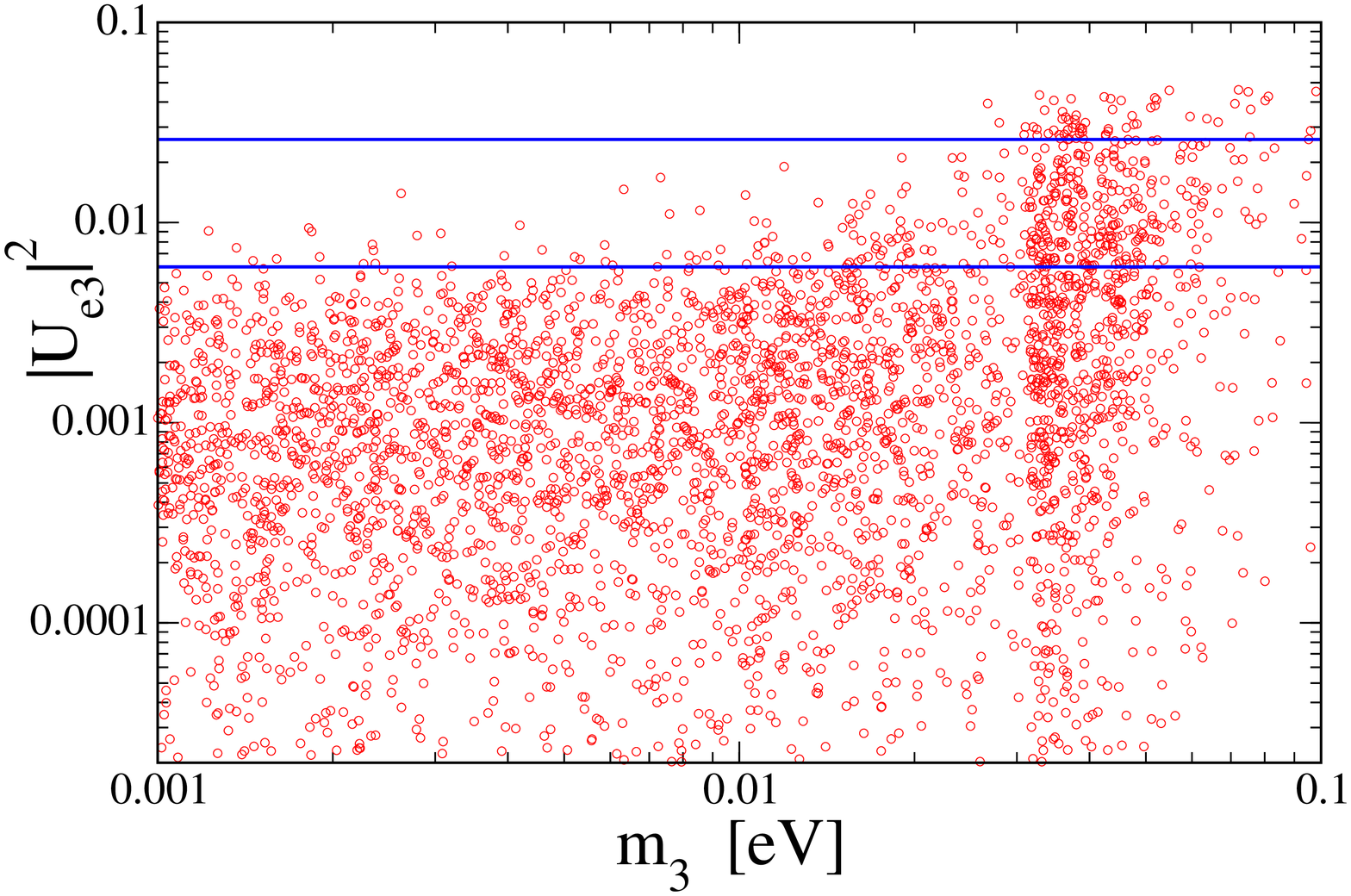,width=3.0in}
}
\caption{\label{fig:AR}Scatter plot 
for an explicitly broken TBM mass matrix 
of $|U_{e3}|$ against the smallest neutrino 
mass for the normal (left) and inverted (right) mass ordering.}
\end{center}
\end{figure}

%
\section{\label{sec:expl}Breaking Tri-bimaximal Mixing explicitly}
%
%
We can explicitly break TBM by perturbing the neutrino 
mass matrix. As for RG effects, we will see that there 
is crucial dependence on the neutrino mass ordering and 
values of neutrino masses.
In its general form, $m_\nu$ leading to TBM is given in Eq.~(\ref{eq:mnutbm}). 
A possible strategy  to perturb TBM, outlined in detail in 
Ref.~\cite{AR}, is to modify the mass matrix in 
the following way: 
\be \label{eq:mnu_dev}
m_\nu =   
\left(
\bad 
A \, (1 + \epsilon_1) & B \, (1 + \epsilon_2) 
& B \, (1 + \epsilon_3)\\[0.2cm]
\cdot & \frac{1}{2} (A + B + D) \, (1 + \epsilon_4) 
& \frac{1}{2} (A + B - D) \, (1 + \epsilon_5)\\[0.2cm]
\cdot & \cdot & \frac{1}{2} (A + B + D) 
\, (1 + \epsilon_6)
\ea 
\right) .
\ee
%
The complex perturbation parameters $\epsilon_i$ are taken to 
be $|\epsilon_i| \le 0.2$ for $i = 1 - 6$ with their 
phases $\phi_i$ allowed to lie between zero and $2\pi$. 
In case of a normal hierarchy, one finds \cite{AR} that 
$|U_{e3}|^2$ is of order $\epsilon^2 \, R$, where 
$\epsilon$ is the magnitude of one of the $\epsilon_i$, and 
$R = \dms/\dma$. 
Hence, a too small value of $|U_{e3}|^2$ is generated in this case. 
It turns out that at least $m_1 \simeq 0.015$ eV is required in 
order to generate $|U_{e3}|^2$ above 0.006. This is illustrated in 
Fig.\ \ref{fig:AR}.

Such values correspond to a scenario with a
partial mass hierarchy: 
$m_1 \simeq m_2 \ls m_3$. With the increase of $m_1$ 
starting from 0.015 eV, 
the maximal value of $|U_{e3}|$ grows almost linearly with $m_1$. 
In contrast, in the case of inverted hierarchy (ordering),  
one can generate large values of $|U_{e3}|$ even for 
a vanishing value of the smallest neutrinos mass $m_3$. 
For quasi-degenerate neutrinos,
obviously, 
sizeable values of $|U_{e3}| \simeq 0.1$ can also be generated.
In addition, in the cases of  neutrino mass spectrum 
with partial hierarchy, with  
inverted hierarchy and of
quasi-degenerate type, 
there exists a correlation between the 
effective Majorana mass in $\betabeta$-decay 
and the value of $|U_{e3}|$ thus generated.  

 To illustrate the above comments, 
consider the following 
analytic estimates in the case of 
spectrum with inverted ordering.   
We first set $\alpha_2 = \pi$. In this case 
one has $A \simeq \sqrt{\dma}/3$ and $B \simeq -2\sqrt{\dma}/3$. 
Consider now a perturbation of the form 
\be 
m_\nu =   
\left(
\bad 
A  & B \, (1 + \epsilon) 
& B \, (1 - \epsilon)\\[0.2cm]
\cdot & \frac{1}{2} (A + B + D) 
& \frac{1}{2} (A + B - D) \\[0.2cm]
\cdot & \cdot & \frac{1}{2} (A + B + D) 
\ea 
\right) .
\ee
%
for real $\epsilon$, either negative or positive. 
In this case we get the largest effects on 
$|U_{e3}|$ and $\sin^2 \theta_{23}$ \cite{AR}: 
\be
|U_{e3}|^2 \simeq  \epsilon^2 \left( \frac{8}{81} 
+ \frac{16}{27} \, \frac{m_3}{\sqrt{\dma}}
\right) \ls 10^{-2} 
\mbox{ and } 
\left|\sin^2 \theta_{23} - \frac 12\right| \simeq \frac 89 \, 
\epsilon^2 \gs 0.18~.
\ee
%
Note that $A \simeq \sqrt{\dma}/3 \simeq 0.016$ eV is the minimal 
possible value of the
$\betabeta$-decay effective Majorana mass 
in the case of spectrum with inverted hierarchy under discussion. 
The cancellation arises due to the chosen 
CP conserving value of the Majorana phase $\alpha_2$. 

In the other extreme case of $\alpha_2 = 0$, 
we have $B/A \simeq \frac 16 \, \dms/\dma$ 
and we find that $|U_{e3}|^2$ is at most 
of order $(\epsilon \, B/A)^2 \simeq 10^{-6}$ 
and therefore completely negligible. 

 In the case of perturbed 
$\mu\mu$ and $\tau\tau$ entries of $m_{\nu}$, 
\be 
m_\nu =   
\left(
\bad 
A  & B 
& B \\[0.2cm]
\cdot & \frac{1}{2} (A + B + D) \, (1 - \epsilon)
& \frac{1}{2} (A + B - D) \\[0.2cm]
\cdot & \cdot & \frac{1}{2} (A + B + D) \, (1 + \epsilon)
\ea 
\right) ,
\ee
%
the largest possible deviation of 
$\theta_{23}$ from $\pi/4$ is obtained for 
$\alpha_2 = 0$:  $|\sin^2 \theta_{23} - \frac 12| \simeq 
\epsilon/2 \simeq 0.1$. 

\begin{figure}[t]
\begin{center}
\parbox{3in}{
\epsfig{file=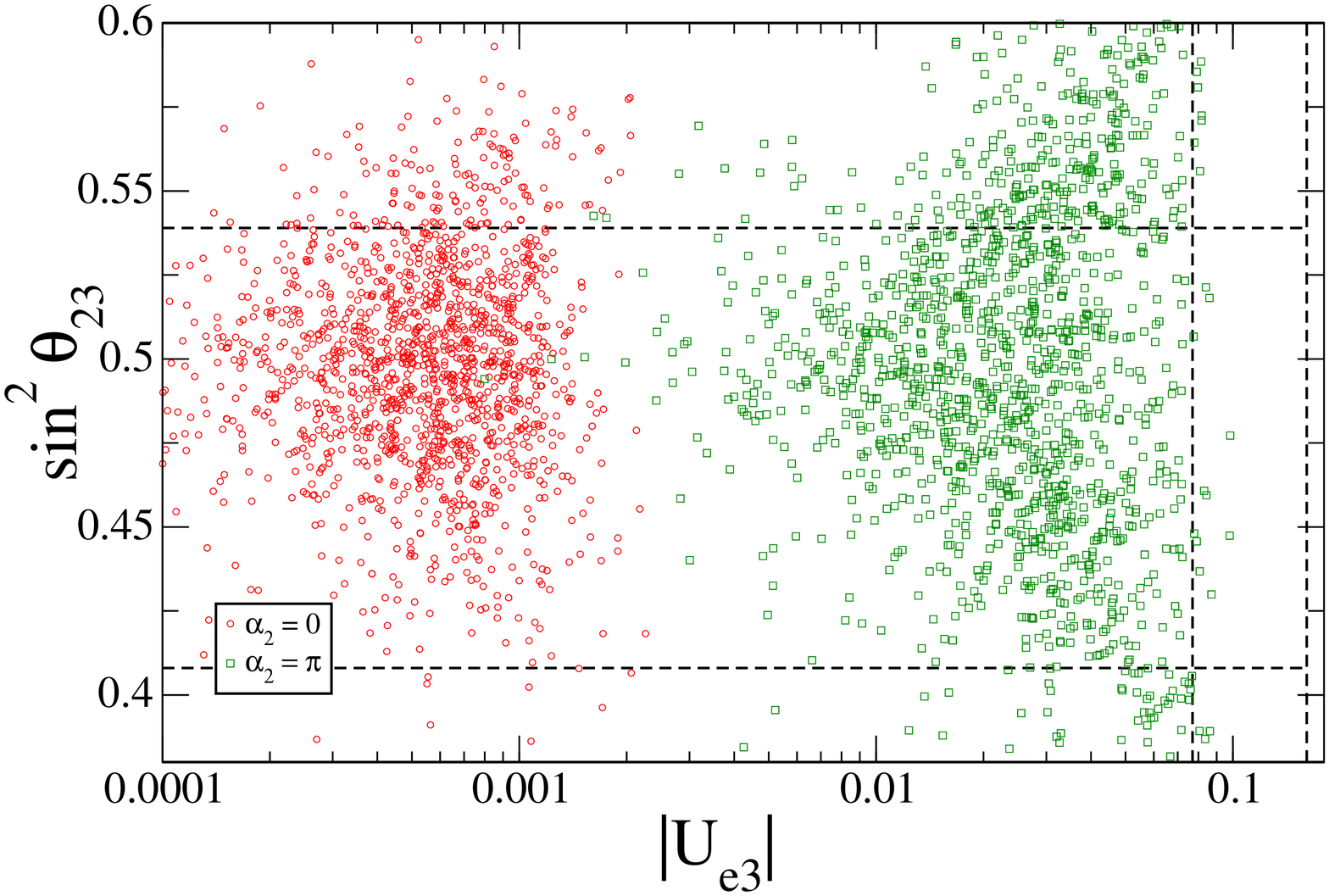,width=3in}
}
\parbox{3in}{
\epsfig{file=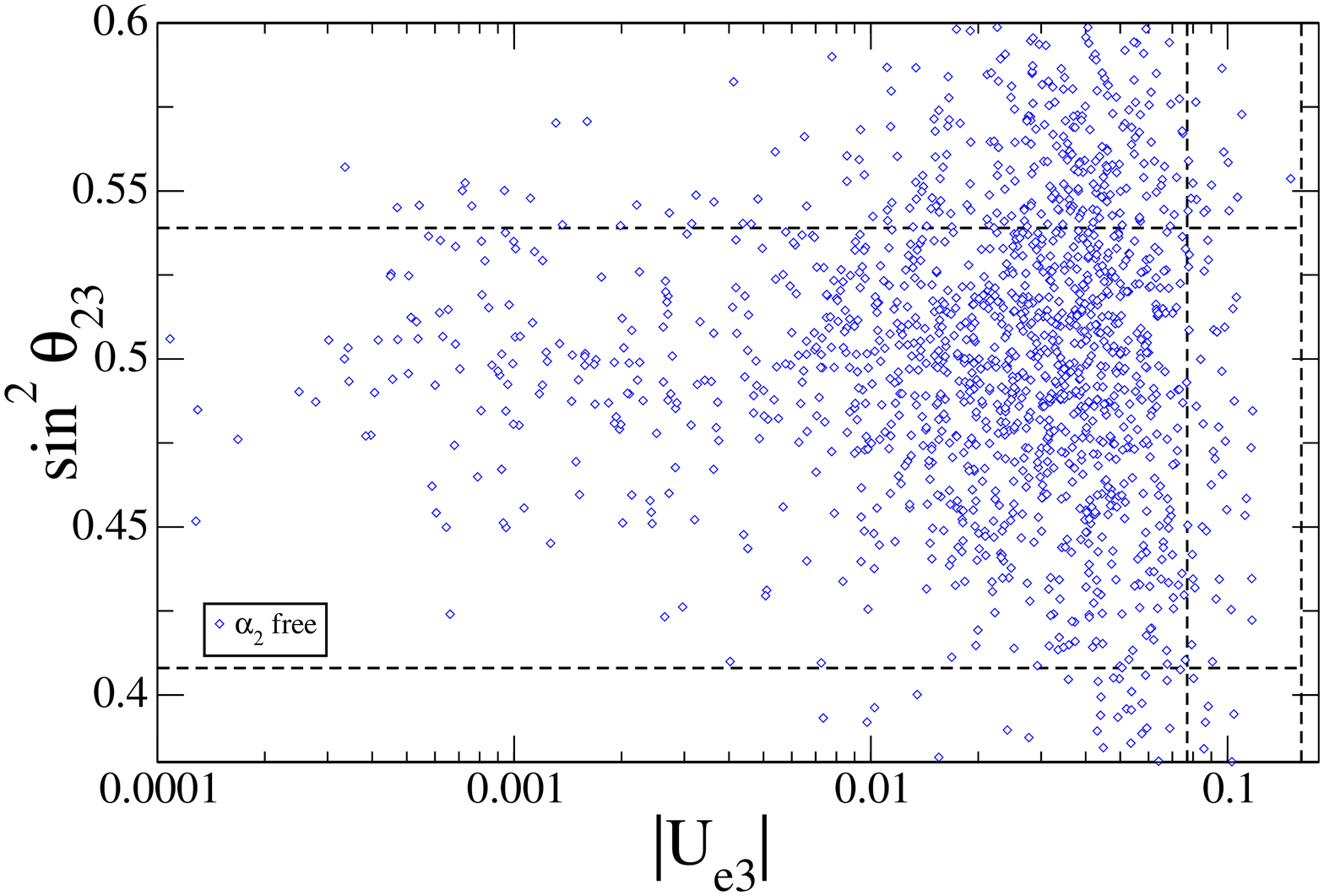,width=3.0in}
}
\parbox{3in}{
\epsfig{file=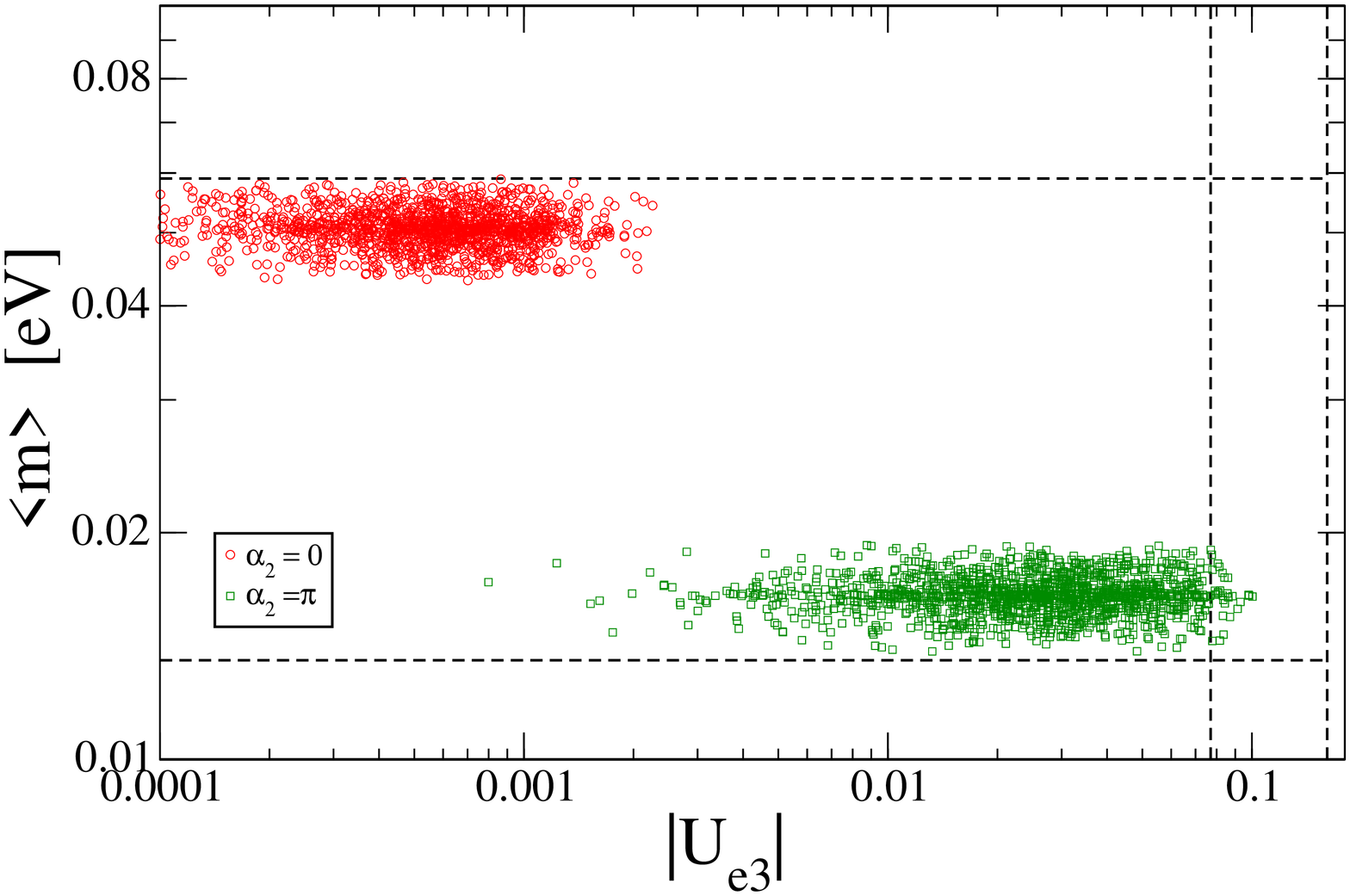,width=3.0in}
}
\parbox{3in}{
\epsfig{file=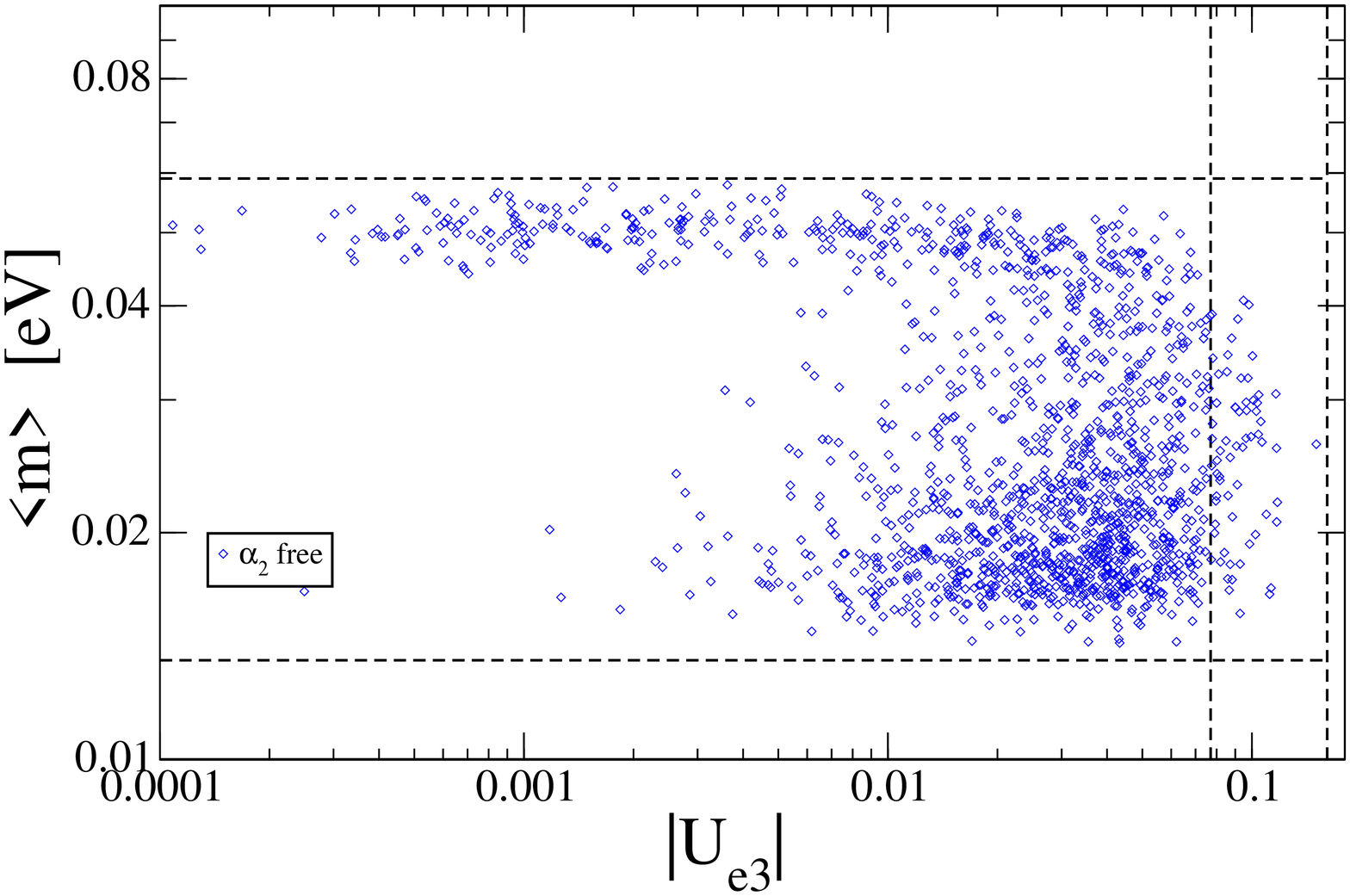,width=3.0in}
}
\caption{\label{fig:IHexpl}Scatter plot of $|U_{e3}|$ against $\sin^2 
\theta_{23}$ as well as of $|U_{e3}|$ against $\meff$ 
for an explicitly broken TBM mass matrix in case of an
inverted hierarchy. The left plots show the cases $\alpha_2 = 0$ and 
$\alpha_2 = \pi$, the right plots have free $\alpha_2$. Indicated are
also the $1\sigma$ ranges of the oscillation parameters, and the upper
and lower limits of the effective mass.}
\end{center}
\end{figure}

We conclude that \cite{AR}, if initially the phase $\alpha_2$ takes
a CP conserving value of $\pi$ and for an inverted hierarchy 
neutrino mass spectrum,  perturbed TBM leads to 
values of the $\betabeta$-decay effective Majorana 
mass close to the minimal one, 
$\meff = c_{13}^2 \, \sqrt{\dma}\cos 2\theta_{12}$. 
These values  
are correlated with sizable values of $|U_{e3}|$ and relatively large 
deviations from maximal atmospheric
neutrino mixing. The benchmark value of 
$|U_{e3}|$ from Eq.~(\ref{eq:bari}) can be reconciled with 
minimal allowed values of the effective Majorana mass. 
In contrast, if the effective Majorana mass $\meff$
is close to its possible maximal value, 
$\meff \simeq c_{13}^2 \, \sqrt{\dma}$, 
negligible values of $|U_{e3}|$ are predicted. 
Hence, the benchmark value of 
$|U_{e3}|$ from Eq.~(\ref{eq:bari}) cannot be reconciled with 
values of $\meff$ close to its maximal value.
The expected deviation from $\sin^2 \theta_{23} = \frac 12$ is also
smaller than in the previous case. 
It turns out, however, that the case of 
free $\alpha_2 \neq 0$ or $\pi$ allows 
non-minimal values of \meff~for sizeable $|U_{e3}|$ as well (see below), 
which means that the correlations discussed above 
rely on extreme initial values of $\alpha_2$. 

\begin{figure}[t]
\begin{center}
\parbox{3in}{
\epsfig{file=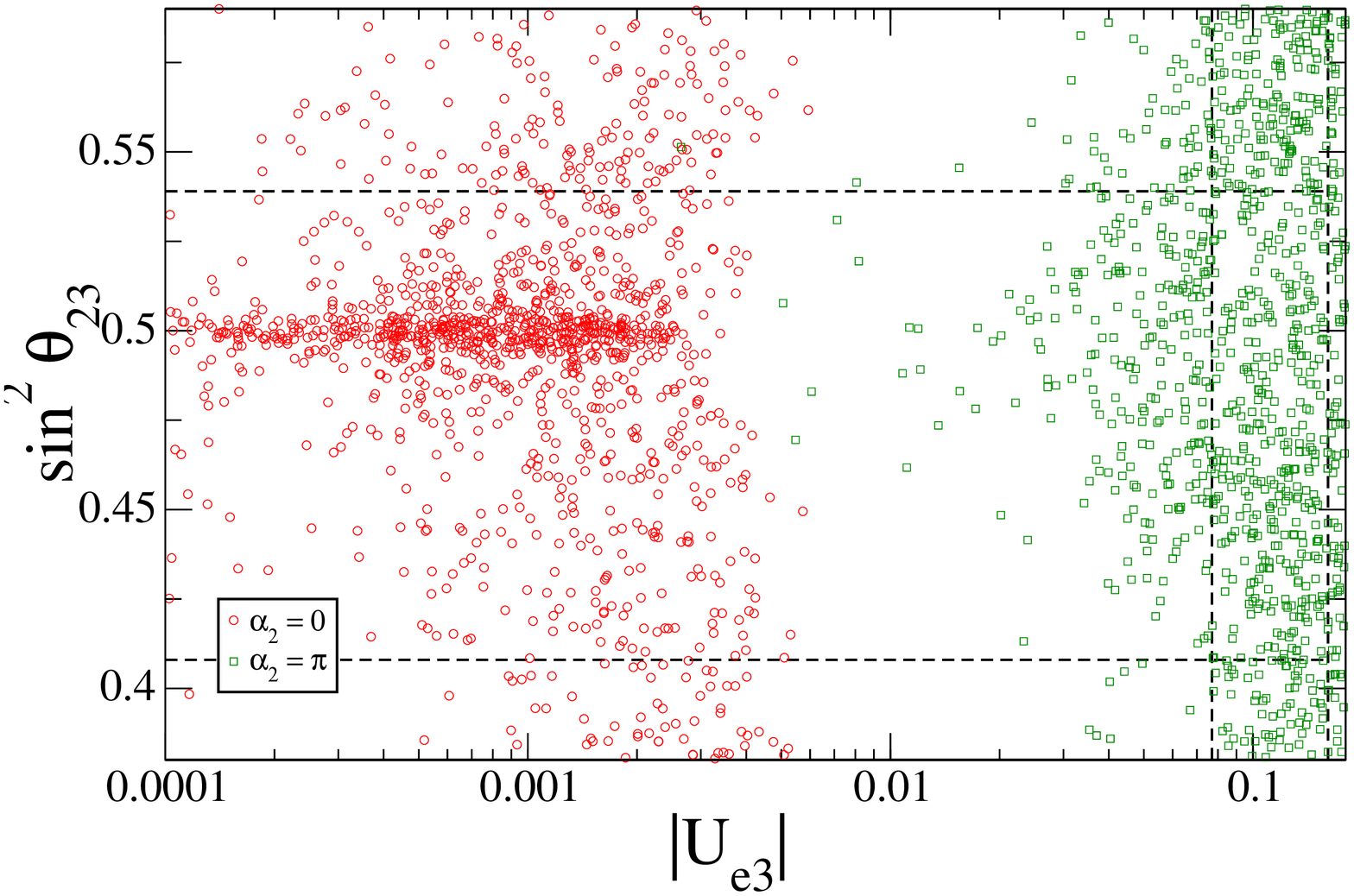,width=3in}
}
\parbox{3in}{
\epsfig{file=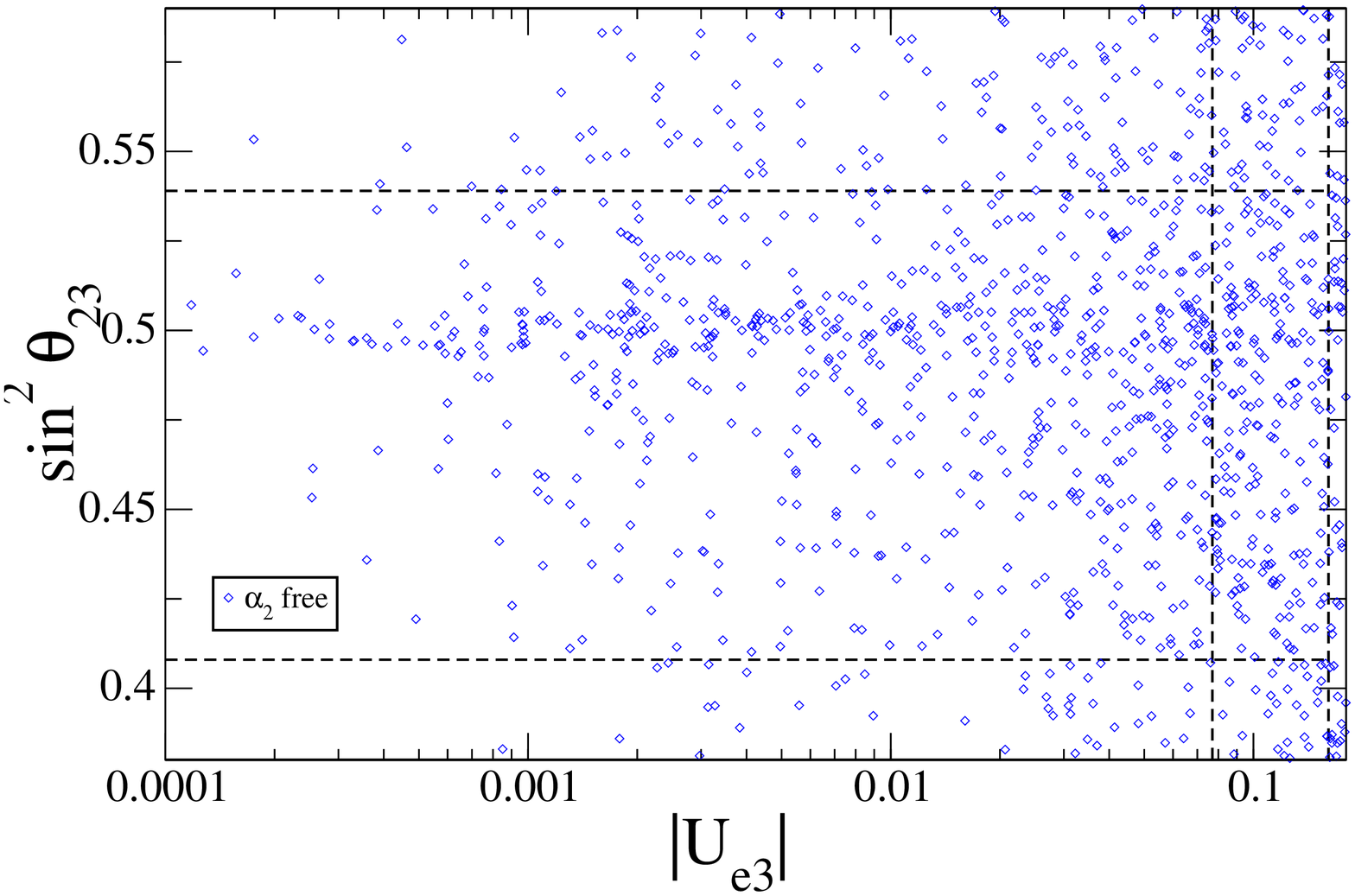,width=3.0in}
}
\parbox{3in}{
\epsfig{file=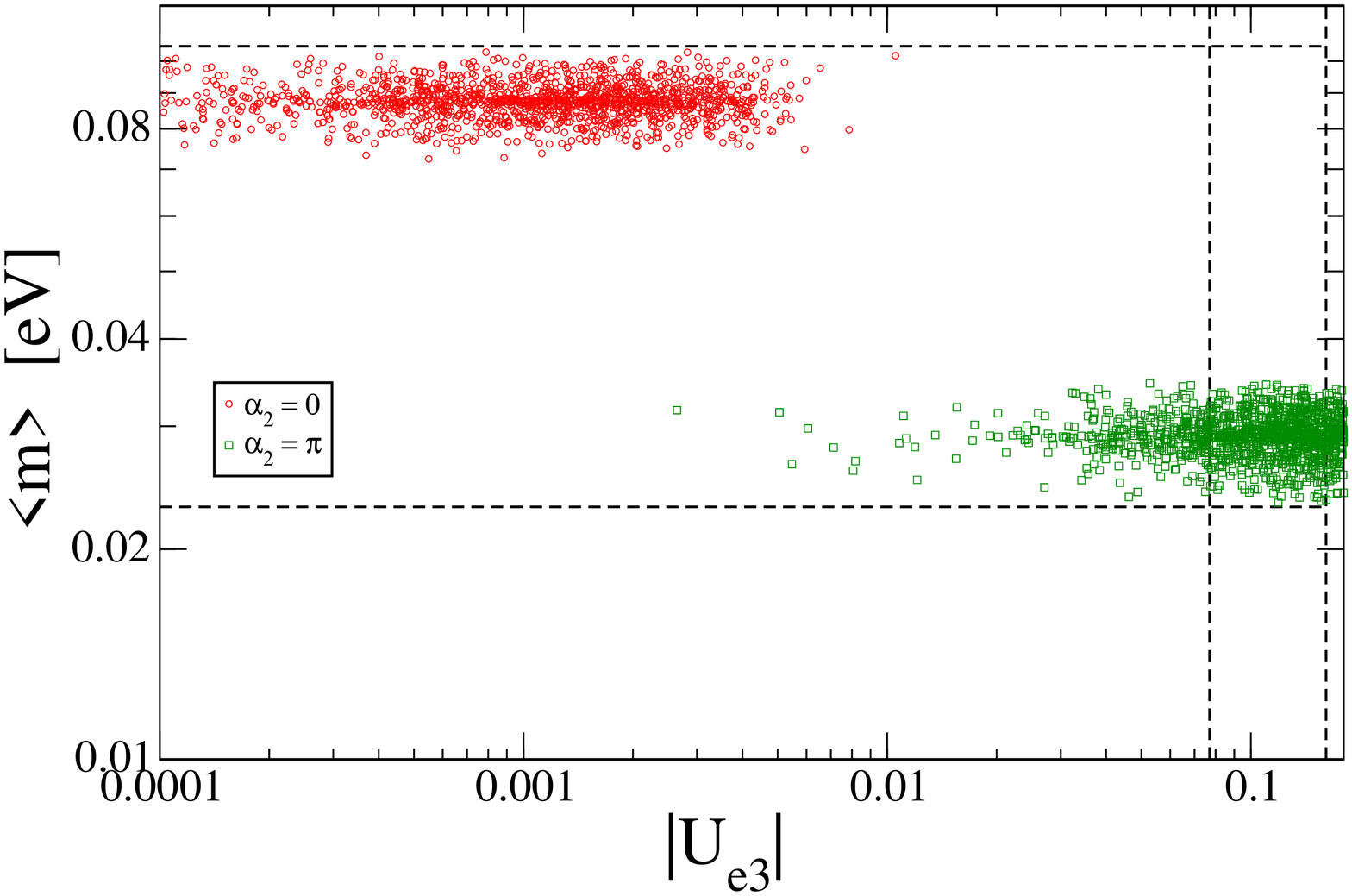,width=3.0in}
}
\parbox{3in}{
\epsfig{file=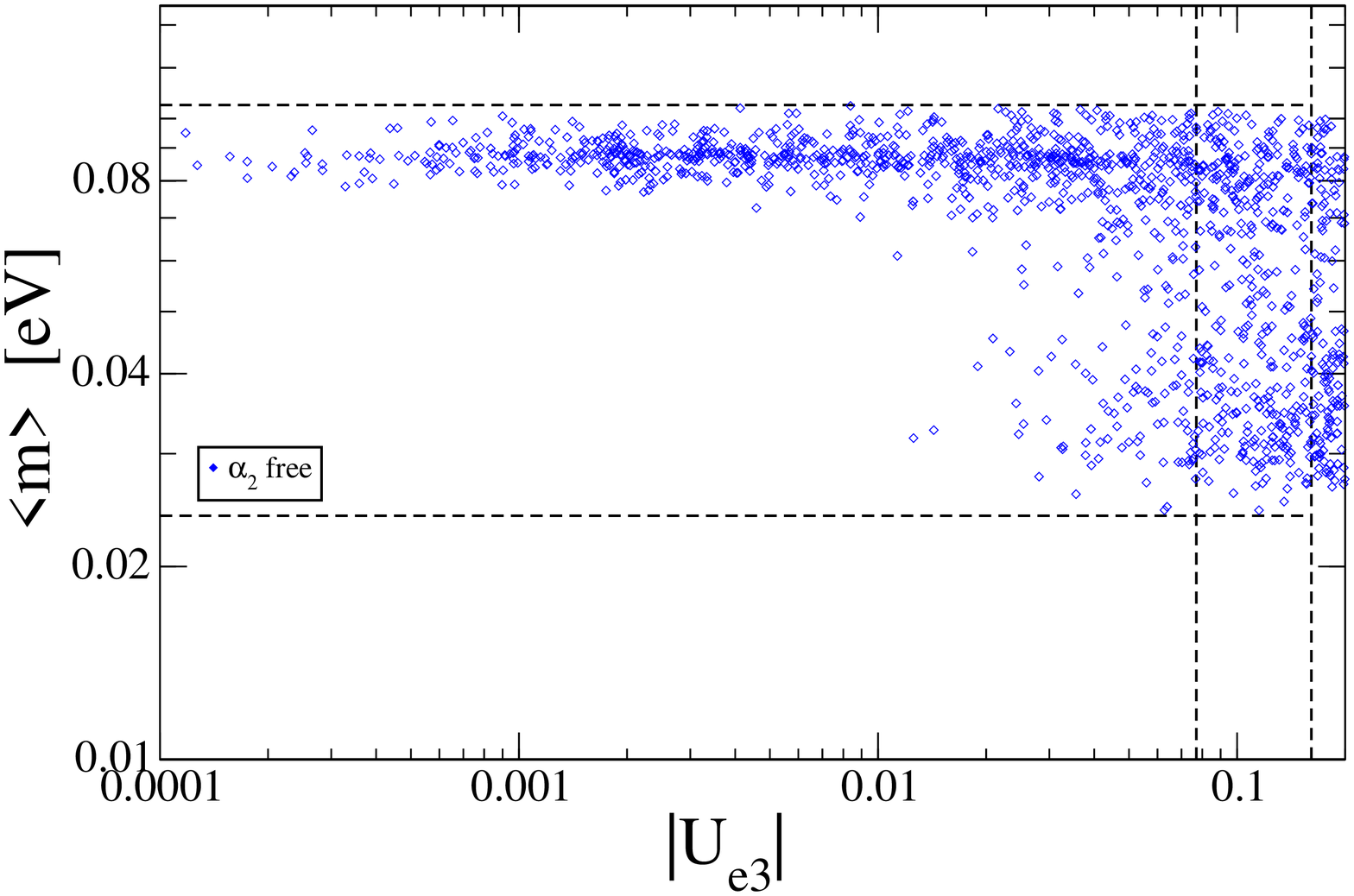,width=3.0in}
}
\caption{\label{fig:QDexpl}Same as Fig.~\ref{fig:IHexpl} for 
quasi-degenerate neutrinos.}
\end{center}
\end{figure}
%
In Fig.~\ref{fig:IHexpl} we show 
scatter plots resulting from a corresponding 
numerical analysis. We have 
diagonalized Eq.~(\ref{eq:mnu_dev}), where we have taken random values
for the complex $\epsilon_i$, by starting with 
$m_2 = 0.051$ eV, $m_1 = 0.0502424$ eV and $m_3 = 0.01$ eV. 
We required the resulting oscillation observables to lie in 
their $3\sigma$ ranges. We have also chosen as initial values 
$\alpha_2 = 0$, $\pi$, but let $\alpha_2$ vary freely as well. 
The largest and smallest possible values of the effective mass 
\meff~are approximately $0.059$ eV and $0.0135$ eV, respectively, 
and we have indicated them in the figure. The analytical estimates
from above are confirmed here. 
Fig.~\ref{fig:QDexpl} shows the same analysis for quasi-degenerate
neutrinos, where we have started with $m_3 = 0.10$ eV, 
$m_2 = 0.08778$ eV and $m_1 = 0.08735$ eV. The effective Majorana 
mass $\meff$ in this case lies 
between $0.023$ eV and $0.105$ eV. 
In both cases it is evident that 
the value of $\theta_{23}$ is not a good discriminator. 
In particular,  maximal atmospheric neutrino mixing is always possible. 

%
\section{\label{sec:concl}Conclusions and Summary}
%
%

Tri-bimaximal mixing provides a very close description of 
neutrino mixing angles. However, the present hint of non-zero 
$\theta_{13}$ coming from analyses of the global neutrino oscillation 
data may indicate that it is broken. 
In this paper we consider three breaking mechanisms from exact TBM 
-- charged lepton corrections, radiative corrections and 
explicit breaking. While the deviation from maximal $\sin^2 
\theta_{23} = \frac 12$ is allowed by the data to be of the 
same order 0.1 as the values of $|U_{e3}|$ that we study, 
the challenge is to simultaneously keep the deviations from $\sin^2 
\theta_{12} = \frac 13$ of order $|U_{e3}|^2$ or below.
 For definiteness we choose in our analysis the $1\sigma$ range given in 
Eq.~(\ref{eq:bari}): $0.077 \le |U_{e3}| \le 0.161$. The main 
results of this work are summarized in Table \ref{tab:sum}.\\ 

We first 
assume CKM-like charged lepton corrections from $U_\ell$ to 
$U_\nu$ corresponding to tri-bimaximal mixing. The correction 
parameter $\lambda$, which is the sine of the 12-rotation in the
usual parametrization of $U_\ell$, can be restricted as 
$0.104 \leq \lambda \leq 0.247$ from the 
current 1$\sigma$ ranges of the mixing angles. We note that the 
sine of the Cabibbo angle is included in this range, but one third of
it is not. 
In this picture $|U_{e3}| \simeq  \lambda/\sqrt{2}$. 
A sizable value of $U_{e3}$ therefore implies a sizable $\lambda$. 
Suppressing the leading (${\cal O}(\lambda)$) correction 
to $\sin^2\theta_{12}$ is possible by choosing the Dirac 
CP phase in neutrino oscillations to be $\pi/2$ or $3\pi/2$
corresponding to maximal CP violation in neutrino oscillations. 
The charged lepton corrections to tri-bimaximal mixing do not depend 
on the neutrino mass values and their ordering. 
The atmospheric neutrino mixing parameter 
$\sin^2 \theta_{23}$ is deviated from $\frac 12 $ by terms of 
order $|U_{e3}|^2$. To be precise, it 
is within the range $0.44 \ls \sin^2 \theta_{23} \ls 0.53$.  
In particular it is allowed to be maximal. 

\begin{table}[t]
\begin{center} 
\begin{tabular}{|c|c|c|c|} \hline 
 & charged leptons & renormalization (MSSM)  & explicit breaking \\ \hline \hline
$ \sin^2 \theta_{23}$ & $0.44 - 0.53 $ & 
$\baz 0.55 - 0.64 & (\dma > 0) \\ 
  0.33 - 0.45 & (\dma < 0)
\ea $ 
& --- \\ \hline 
$ |U_{e3}|$ & $\simeq \frac{\D \lambda}{\D \sqrt{2}} $ & 
$\propto \frac{\D m_0^2}{\dma} \, (1 + \tan^2 \beta)$
& $\baz \propto \epsilon & \rm (IH) \\ 
 \propto \epsilon \, m_1/\sqrt{\Delta m^2_{\rm A}} & \rm (PD/QD) \ea $ \\ \hline 
 mass & --- & QD: $m_0 \, \tan \beta \simeq (4 - 7)$ eV 
& IH, PD, QD\\ \hline 
\meff & --- & $m_0 \, c_{13}^2 \, \cos 2 \theta_{12} $ & 
$\baz m_0 \, c_{13}^2 \, \cos 2 \theta_{12} & \rm (QD) \\ 
\sqrt{\dma} \, c_{13}^2 \, \cos 2 \theta_{12} & \rm (IH) \\
\ea $ \\ \hline 
CP & $\ba \mbox{oscillations: almost} \\ \mbox{maximal CP violation} \ea $ 
& $\alpha_2 \simeq \pi$ & 
$ \ba \mbox{large $|U_{e3}|$ requires} \\ 
\mbox{suppressed \meff~only} \\ 
\mbox{when initially $\alpha_2 \simeq \pi$} \ea $ \\ \hline 
\end{tabular}
\caption{\label{tab:sum}Requirements on and predictions of 
the three breaking scenarios in order to 
generate the $1\sigma$ range 
$0.077 \le |U_{e3}| \le 0.161$.  
IH denotes inverted hierarchy, while PD stands for a 
partial hierarchal and QD for a quasi-degenerate mass scheme.}
\end{center}
\end{table}

Generating a large $|U_{e3}|$ via radiative corrections 
implies quasi-degenerate neutrinos, namely $m_0 \gs 2.6$ eV for the SM 
when $U_{e3} = 0$ at the high scale. Thus, the current neutrino mass 
limits rule out a possible RG origin of 
$|U_{e3}| \simeq 0.1$ in the SM. 
For the MSSM one requires 
$m_0 \gs 0.8 \, (0.2)$ eV with $\tan\beta = 5\,(20)$. The implied
constraints of the $1\sigma$ range of $|U_{e3}|$ 
on $m_0$ and $\tan \beta$ can be summarized as 
$4 \ls (m_0/{\rm eV}) \tan\beta \ls 7$. 
The running in the MSSM predicts that $\sin^2 \theta_{12}$ increases
from its initial high scale value. 
Large running of $\theta_{13}$ to generate $|U_{e3}| \simeq 0.1$
together with the requirement that $\sin^2\theta_{12}$ is within its current 
3$\sigma$ range forces the Majorana phase $\alpha_2 \simeq \pi$. 
Interesting correlations are also obtained between $\sin^2\theta_{13}$ 
and $\sin^2\theta_{23 }$. The latter parameter is necessarily
non-maximal and lies in the range 
$0.55 \le \sin^2 \theta_{23} \le 0.64$ for a normal ordering and 
$0.33 \le \sin^2 \theta_{23} \le 0.45$ for an inverted ordering. 
The RG evolved effective neutrino Majorana 
mass observed in neutrino-less double beta decay 
is found to lie close to its minimum allowed value 
because of the constraint of $\alpha_2 \simeq \pi$.
In case of $\tan \beta = 5$, \meff~lies between 
0.26 and 0.50 eV, to be compared with its general upper and lower 
limits of 0.2 eV and 1.4 eV. If $\tan \beta = 20$, then 
$0.07~{\rm eV}\ls \meff \ls 0.11~{\rm eV}$, 
while in general the effective mass could be 
in between 0.05 eV and 0.34 eV.  

We also consider the possibility of deviating from tri-bimaximal mixing 
by adding explicit breaking terms to the neutrino mass matrix, i.e., 
every entry is multiplied with an individual factor $1 + \epsilon_i$. 
For this mechanism to generate sizable $|U_{e3}|$ the neutrino 
mass spectrum has to be partially degenerate,  or quasi-degenerate, or
with inverted hierarchy.  Atmospheric neutrino mixing is 
allowed to take any of its currently allowed values, including 
$\sin^2 \theta_{23} = \frac 12$. 
In this breaking scenario the requisite sizeable $|U_{e3}|$ 
value cannot be reconciled with initial 
maximal values of the effective Majorana mass 
governing neutrino-less double 
beta decay, corresponding to the three indicated 
types of neutrino mass spectrum.\\

To sum up, the CP violating phases in the 
neutrino mixing matrix play a crucial role 
for having only relatively small corrections to $\theta_{12}$ 
when large corrections to $U_{e3} = 0$ are generated.  
This interesting fact together with the 
predictions for $\theta_{23}$ may be used to distinguish breaking 
scenarios to tri-bimaximal mixing.

%
\vspace{0.3cm}
\begin{center}
{\bf Acknowledgments}
\end{center}
%
%
The authors would like to thank Amol Dighe for useful discussions. 
The work of W.R.~was supported by the ERC under the Starting Grant 
MANITOP and by the Deutsche Forschungsgemeinschaft 
in the Transregio 27. 
S.R.~was partially supported by the 
Max Planck--India partner group project between 
Tata Institute of Fundamental Research
and the Max--Planck Institute for Physics. 
S.G.~acknowledges support from the Neutrino Project under the 
XI-th plan of Harish--Chandra Research Institute.
This work was supported also in part by the Italian INFN under 
the program ``Fisica Astroparticellare'' (S.T.P.).

\end{document}